\documentclass[12pt]{article}
\pdfoutput=1
\usepackage{sectsty}
\usepackage{mathrsfs}
\allsectionsfont{\sffamily}
\newcommand{\5}{$_5$}

\usepackage[nosort]{cite}

\usepackage[pdftex]{graphicx}
\usepackage[font={scriptsize,singlespacing}]{caption}
\usepackage{epstopdf}

\usepackage{amsmath}
\usepackage{amssymb}
\usepackage{subfigure}
\usepackage{hyperref}
\usepackage{url}
\usepackage{xcolor}
\usepackage{color}
\definecolor{amaranth}{rgb}{0.9, 0.17, 0.31}
\definecolor{purple(munsell)}{rgb}{0.62, 0.0, 0.77}
\definecolor{americanrose}{rgb}{1.0, 0.01, 0.24}
\definecolor{palatinateblue}{rgb}{0.15, 0.23, 0.89}
\definecolor{royalblue(web)}{rgb}{0.25, 0.41, 0.88}
\definecolor{hanpurple}{rgb}{0.32, 0.09, 0.98}
\definecolor{beaublue}{rgb}{0.74, 0.83, 0.9}
\definecolor{carminered}{rgb}{1.0, 0.0, 0.22}
\definecolor{brightpink}{rgb}{1.0, 0.0, 0.5}
\definecolor{vividviolet}{rgb}{0.62, 0.0, 1.0}
\hypersetup{ linktoc=all,
    colorlinks, linkcolor={palatinateblue},
    citecolor={brightpink}, urlcolor={amaranth}}

\newcommand{\changeurlcolor}[1]{\hypersetup{urlcolor=#1}}    
\def\sideremark#1{\ifvmode\leavevmode\fi\vadjust{\vbox to0pt{\vss% the remark
 \hbox to 0pt{\hskip\hsize\hskip1em%                          will appear only
 \vbox{\hsize2cm\tiny\raggedright\pretolerance10000%          on the side
 \noindent #1\hfill}\hss}\vbox to8pt{\vfil}\vss}}}%
\newcommand{\bo}{\raise-1mm\hbox{\Large$\Box$}}

\newcommand{\f}[2]{\frac{#1}{#2}}

\newcommand{\la}{\langle}
\newcommand{\ra}{\rangle}
\newcommand{\w}{\omega}
\newcommand{\kp}{\kappa}
\newcommand{\be}{\begin{equation}}
\newcommand{\ee}{\end{equation}}
\newcommand{\bea}{\begin{eqnarray}}
\newcommand{\eea}{\end{eqnarray}}

\renewcommand{\d}[1]{\ensuremath{\operatorname{d}\!{#1}}}

\begin{document}
\thispagestyle{empty}
\begin{center}

\null \vskip-1truecm \vskip2truecm

{\LARGE{\bf \textsf{On Horizonless Temperature with an Accelerating Mirror}}}
%\vskip0.1truecm
%{\Large{\bf \textsf{A Finite Energy Moving Mirror\\}}}
%{\LARGE{\bf \textsf{On Particle Production and Energy Emission \\}}}

\vskip1truecm
\textbf{\textsf{Michael R.R. Good and Khalykbek Yelshibekov}}\\
{\footnotesize\textsf{Department of Physics, School of Science and Technology, Nazarbayev University, \\53 Kabanbay Batyr Ave., 
Astana, 010000, Republic of Kazakhstan}\\
{\tt Email: michael.good@nu.edu.kz, khalykbek.yelshibekov@nu.edu.kz}}\\

\vskip0.4truecm
\textbf{\textsf{Yen Chin Ong}}\\
{\footnotesize\textsf{(1) Center for Astronomy and Astrophysics, Department of Physics and Astronomy,\\
Shanghai Jiao Tong University, Shanghai 200240, China\footnote{Present address.}}}\\
{\footnotesize\textsf{(2) Nordita, KTH Royal Institute of Technology and Stockholm University, \\ Roslagstullsbacken 23,
SE-106 91 Stockholm, Sweden}\\ \vskip0.05truecm
{\tt Email: ongyenchin@sjtu.edu.cn}}\\

%\vskip0.4truecm
%\textbf{\textsf{Paul Anderson}}\\
%{\footnotesize \textsf{Department of Physics, \\ Wake Forest University, \\ Winston-Salem, North Carolina, USA}\\
%{\tt Email: anderson@wfu.edu}}\\ 

%\edz{\color{red} Should we follow the HEP convention to arrange authors by alphabetical order? Of couse this is not necessary and I am fine with the current one as well.\color{blue}{That might be a good idea.  I have a preference for as it is, but let's ask Paul what he thinks.}}

\end{center}
\vskip1truecm \centerline{\textsf{ABSTRACT}} \baselineskip=15pt

\medskip
%While black hole evaporation can exhibit puzzling features such as negative entanglement entropy and negative energy flux emission, such phenomena are present in flat spacetime moving mirror models. In this work, to facilitate a further understanding of these puzzles, we construct a unitary moving mirror trajectory whose time evolution only emits a finite amount of energy, and lacks an acceleration horizon.  In the limit that mirror drifts to the speed of light, the model is that of null shell collapse to form a black hole.  We study the energy flux, particle flux and entanglement entropy production to the right, as well as the left, sides of the mirror. For speeds less than light, the entanglement entropy tends to a constant at late times according to the right observers, demonstrating that its black hole correspondence, if it exists, ends as a black hole remnant when evaporation eventually stops.  

A new solution of a unitary moving mirror is found to produce finite energy and emit thermal radiation despite the absence of an acceleration horizon.  In the limit that the mirror approaches the speed of light, the model corresponds to a black hole formed from the collapse of a null shell.  For speeds less than light, the black hole correspondence, if it exists, is that of a remnant.

\vskip0.4truecm
\hrule
%\end{abstract}
%\pacs{03.70.+k, 04.62.+v, 04.60.-m} %04.62.+v is quantum fields in CS, and 03.70.+k is quantized fields, 04.60.-m is quantum gravity
%\keywords{dynamical Casimir effect, moving mirrors, negative energy flux}%Use showkeys class option if keyword
%\maketitle
%\tableofcontents

\section{Introduction: Some Puzzles with Moving Mirrors and Evaporating Black Holes}

The discovery of Hawking radiation from a black hole came at a surprise over 40 years ago \cite{Hawking2}, since it implies that given enough time, black holes in asymptotically flat spacetimes would radiate off their mass; and if no new physics comes into play, they would eventually evaporate completely. This leads to the information paradox \cite{Hawking3} --- where did the information that falls into a black hole disappear to, if the black hole disappears completely? Attempts to recover information from evaporating black holes continue to produce new paradoxes, such as the firewall controversy \cite{amps, apologia, sam}, which threatens the conventional understanding that there should be nothing unusual --- much less a diverging energy density --- at the low curvature spacetime region near the event horizon of a sufficiently massive black hole. In fact, the situation is worse than previously thought: if a firewall does exist, then in principle it can be much further away from the event horizon \cite{naked}, so that an unexpected space traveler would hit it and be burned to death even if he is nowhere near a black hole. While the resolution of the information paradox may require us to fully understand how to unify general relativity with quantum field theory, it is certainly possible that progress can nevertheless be made without a working theory of quantum gravity\footnote{Though we should probably first agree on what counts as a resolution of the paradox. See \cite{huh}.}. In recent years, it has become increasingly clear that we at least need to understand the subtle ``physics of information'' \cite{Landauer, Adami, Wolf}, and how it applies to black holes \cite{0311049, 0708.4025, 0808.2096,1111.6580,1301.4504,1301.4505,1402.5674}, in order to understand the information retrieval process from Hawking quanta. 

In fact, a central piece of the puzzle regarding information loss, is the understanding of entanglement entropy, $S$. Let us consider the formation of a black hole from a gravitational collapse of some matter in a pure state. It is often believed that the entanglement entropy of the Hawking radiation received at null infinity, which should be zero at the beginning before any radiation arrives, should first increase, but then decrease at some point (known as the ``Page time''), so that eventually the entanglement entropy vanishes. Such a ``Page curve'' \cite{page1, page1b, page2} --- the plot of the entanglement entropy against time --- is crucial, since it gives insight into how information may be retrieved from the highly scrambled Hawking emission. There are, however, subtleties that are often overlooked. Notably, any calculation of entanglement entropy necessitates regularizing ultraviolet divergences. One notices that imposing a cutoff is a tricky procedure since modes that have sufficiently high energy at some point in the spacetime can be redshifted at some other point due to spacetime curvature. This implies that a mode that is beyond a cutoff scale can be red-shifted below the scale, so the cutoff is not well-defined \cite{0901.3156, 9403137}. Furthermore, the results obtained could depend on the cutoff scheme. Progress has been made recently with the introduction of the ``causal-splitting regularization'' scheme of Bianchi and Smerlak \cite{Bianchi:2014vea, Bianchi:2014qua, 1409.0144}, which allows one to compute the production of entanglement entropy in a \emph{cutoff-independent} manner. However, even then, there are still a few puzzles regarding the entanglement entropy of an evaporating black hole. We list two such puzzles below.\newline 

\underline{(1) Negative Energy Flux and Negative Entropy:} Firstly, it has been observed, e.g. by Bianchi and Smerlak \cite{Bianchi:2014vea, Bianchi:2014qua}, that mass loss of a black hole, assuming unitarity, is not monotonic. In other words, at some point in time, the mass of an ``evaporating'' black hole actually \emph{increases}. For an observer at infinity, Hawking radiation reduces the mass of a black hole by emitting particles. This means that an asymptotic observer sees a flux of positive energy coming out from the black hole. If the mass increases, albeit briefly\footnote{Abdolrahimi and Page have also shown that for asymptotically flat Schwarzschild black holes, this increase in the mass is barely noticeable.\cite{AP}}, during the course of Hawking evaporation, this would correspond to the emission of negative energy flux (hereinafter, ``NEF'') from the black hole. (NEF emission from evaporating black holes, at least in the case of a (1+1)-dimensional dilaton gravity, was already known in the literature for over 20 years \cite{9508027}.) 
Curiously, an observer equipped only with a particle detector would not be able to see any sign in the spectrum of the Hawking quanta to know that NEF has been emitted \cite{Good:2015nja}. Even more surprising than the emission of NEF, in several models of evaporating black holes, Bianchi and Smerlak have explicitly shown that the \emph{entanglement entropy obtained from their causal-splitting scheme can also become negative} at late times in the course of the evaporation.  The physical interpretation of negative entropy is dependent on the particular type of entropy.  Their physical relationships are subtle and not fully clear: in the literature it has been interpreted as either the radiation being less correlated than the vacuum \cite{1409.0144}, or the radiation being more correlated than the vacuum \cite{9403108v1}.  \newline

\underline{(2) Unitarity and Information Recovery:} Secondly, even with the fully covariant cutoff-independent regularization of Bianchi and Smerlak, it was shown that \cite{Good:2015nja} their criterion only requires that the entanglement entropy tends to a constant asymptotically (both in the far past and in the late future), instead of a more stringent requirement that the entropy should tend to zero so as to recover the pure state. In fact, it is not so surprising that the entanglement entropy can increase monotonically.
Even with the causal-splitting regularization, \emph{as long as there is a cutoff}, then modes can be redshifted by spacetime curvature, and consequently a mode that is below the cutoff scale at future null infinity at some late time $u$, when traced back to the past null infinity, could be well-above the cutoff. Since there is no longer a one-to-one map between early modes and late time modes, it is not surprising that one does not obtain $S(u) \to 0$ at late times. Note that such a phenomenon can be shown to occur even in a simplified model involving a moving mirror in flat spacetime --- as was explicitly shown in \cite{9403137, Good:2015nja} --- in which the quantum field theory is unitary. It was suggested in \cite{Good:2015nja} that the corresponding black hole picture might be a black hole remnant \cite{aharonov, Chen:2014jwq}, and that unitarity is maintained in the sense that if one takes into account both the exterior and the interior of a black hole, then the entire quantum state is pure at all times. In such a picture, information remains hidden inside the ever-shrinking black hole horizon (the interior spacetime can still have a large volume \cite{0901.3156, Chen:2014jwq, 1411.2854, Bhaumik:2016sav, 1503.08245, 1602.04395, 1604.07222}, see however \cite{1510.02182}) and the radiation is never purified. Since the end state is a remnant, the information inside is never destroyed\footnote{Note that there are two types of black hole remnant in the literature: the ``long-lived'' or ``meta-stable'' remnant, and the ``eternal'' remnant \cite{Chen:2014jwq} . The former has lifetime much longer than that of a black hole, that is, proportional to $M^n$ where $n>3$. An eternal remnant, on the other hand, lives forever. Our model corresponds to an eternal red-shifted scenario.}. It would be interesting to investigate if $S(u) \to \text{const.}\neq 0$ at late times \emph{necessarily} implies that the corresponding black hole model ends as a remnant. A main question here, however, is the following: do we understand entanglement entropy enough to base the debate of information loss paradox of black holes on it? The subtleties here suggest that we should be more careful in dealing with this issue.\newline

While Hawking evaporation has been a mainstay of the ``quantum fields under external conditions'' enterprise, the moving mirror model \cite{ Davies:1976hi, Davies:1977yv} has flourished mostly as an ancillary effect.  However, we feel that it is prudent to first understand a moving mirror model in which both issues (1) and (2) can be investigated, after all, a (1+1)-dimensional flat spacetime with a mirror trajectory is much simpler than a (3+1)-dimensional black hole spacetime with non-trivial curvature\footnote{Einstein's gravity is topological in (1+1)-dimensions, but with a suitable coupling to matter fields, gravity need not be trivial in (1+1)-dimensions. This allows one to study black hole evaporation. In our moving mirror model, of course, there is no gravity, only an accelerating mirror, so there is no complication either.}. A trajectory of a moving mirror in spacetime is a reflecting boundary on which  field modes are constrained. A moving mirror excites the modes, thereby producing particles.  The spectrum of the emission depends on the exact trajectory of the mirror. With a suitably chosen trajectory, we can therefore mimic the particle production from, say, a collapsing star. (Readers who are not familiar with moving mirrors can refer to \cite{D0} for a pedagogical exposition.)

Despite many decades of research, physical interpretations of external-potential type problems, like the moving mirrors, have never been made entirely clear.  Moreover, serious confrontations with questions that are evaded in these non-gravitational analogs become inevitable during the study of gravitational problems.  In other words, we would like to emphasize that one should not hope that a moving mirror model can fully resolve information paradox of a \emph{bona fide} black hole, but understanding the subtleties of quantum field theory in a moving mirror model \emph{is} a first step toward the more complicated physics of black holes. 

We have come a long way in the study of moving mirrors. Notably,
Moore\cite{moore}, DeWitt \cite{DeWitt:1975ys}, and later on, Davies and Fulling  \cite{ Davies:1976hi, Davies:1977yv} initiated a program using field theories with external conditions, which eventually demonstrated that quantities like the expectation values of the stress-energy tensor, and the localization of particles using wave packets, can be calculated in various physical problems and used to extract significant physical consequences of the quantum fields. Indeed, there has been renewed interest \cite{Bianchi:2014vea, Bianchi:2014qua, Silva:2016wgo, Stargen:2016euf, Wang:2015axa, Hotta:2015yla, Yeh:2013mca, Nicolaevici:2014ela, Silva:2011fq, Hotta:2015huj, Haro:2008zza, Alves:2010zza, Good:2012cp, Chen:2015bcg}
in the moving mirror model in recent years due, in part, to claims of experimental verification of the dynamical Casimir effect \cite{Wilson:2011, Lahteenmaki:2013mda}. %Rego:2014wta}. 
In many of these particle production scenarios, various systems that exploit the simple mathematical set up of the (1+1)-dimensional moving mirror model have led to novel experimental designs. (See \cite{Chen:2015bcg} for one of the latest proposals.)

%We are motivated to extend this effort in order to better understand the physical origin of the thermal and non-thermal properties underlying both the external-potential type and curved-spacetime type systems. 

As previously mentioned, a central advantage to the moving mirror model is its simplicity.  This is both in general, and in the context of the recent one-to-one correspondence with a black hole \cite{Good:2016oey, paper1, paper2, Good:2016bsq}, which found that for one concrete example, the particle production is exactly the same in both the mirror and black hole cases in (1+1)-dimensions. The simplicity of (1+1)-dimensions allows the crux of Hawking radiation to stand out more clearly, separated from the specialized details associated with higher dimensional curved geometry and back-scattering. With the one-to-one correspondence, the moving mirror model can be treated as an even more precise analogy to Hawking's original argument, and therefore it is of interest to extend the one-to-one correspondence to more physically realistic circumstances, while holding on to this fortunate simplicity.  However, even as a relatively simple theoretical model of black hole evaporation in (1+1) -dimensions, the moving mirror model, has in practice, been very hard to extend to solutions for exact trajectories where the global Bogoliubov coefficients may be evaluated. Few solutions have been found\footnote{For example: the  case of uniform acceleration of Davies-Fulling \cite{Davies:1977yv}; and the case of eternal thermal emission of Carlitz-Willey \cite{Carlitz:1986nh}.} and finite-nonzero-energy cases are scarce.\footnote{See the first known solution found by Walker-Davies \cite{Walker:1982} and the asymptotically static case in Good-Anderson-Evans \cite{Good:2013lca} and a drifting case in Good-Ong \cite{Good:2015nja}.} Nevertheless, mirror trajectories that produce finite amount of energy are precisely those that are physically more realistic. We are therefore interested in such trajectories. 

Although there is pedagogical value associated with the reduced complexity of the direct and straightforward calculations in the moving mirror model in (1+1)-dimensions, one should note that there is a peculiarity in (1+1)-dimensions, namely that Minkowski spacetime in (1+1)-dimensions have \emph{two} sets of past and future null infinities. The right past infinity will be denoted by $\mathscr{I}_R^-$, and the right future infinity $\mathscr{I}_R^+$. Similarly, we also have $\mathscr{I}_L^-$ and $\mathscr{I}_L^+$.
That is to say, its Penrose diagram is a diamond, whereas in higher dimensions, the Penrose diagram for Minkowski spacetime is a triangle\footnote{It is true that one often draws the Penrose diagram of Minkowski spacetime as a diamond even in higher dimensions, but the point is that in (2+1)-dimensions and above, the two points that represent spacelike infinity $i^0$, are actually \emph{one and the same}. This is clearer if one looks at how the conformal diagram wraps around the Einstein cylinder (see Fig.14 on page 122 of \cite{hawkingellis}). In (1+1)-dimensions, the two points are genuinely different infinities.}. The implication is this: if we want to understand quantum field theory of a moving mirror in (1+1)-dimensions, we should also take into account \emph{the left side of the mirror}. 

Yet another motivation to consider both sides of the mirror is the following: In (1+1)-dimensional conformal field theory, there are left and right ``temperatures'', $T_L$ and $T_R$, which are related to the amount of left and right-moving excitations in the field. In a thermal ensemble of states, the thermodynamic temperature is related to these left and right temperatures. The Cardy formula for microstate degeneracy \cite{cardy},
\begin{equation}
S_{\text{micro}}=\frac{\pi^2}{3}(c_L T_L + c_R T_R),
\end{equation}
explicitly involves these temperatures, along with the left and right central charges $c_L$ and $c_R$. 
So from this point of view of thermodynamics, it is natural to include both left and right-moving excitations. 

The most important reason to consider both sides of the mirror, however, is related to the eventual aim to understand information loss of black holes. To do so, as we have argued, a crucial first step is to understand how entanglement works in the simpler case of a moving mirror. In particular, as far as unitarity is concerned, it is important that we include the entire spacetime, this is to ensure that information is not hiding in some part of spacetime that one might otherwise overlook. 

In this work, we are mainly motivated in asking the following questions: 
\begin{quote}
Is there a moving mirror in (1+1)-dimensions, satisfying unitarity in the sense allowed by the Bianchi-Smerlak criterion (namely, $S(u) \to \text{const.}$ as $u \to \pm\infty$), that has no acceleration horizon, produces finite amount of energy, and serves as an analog for Eddington-Finkelstein coordinate null-shell gravitational collapse in its limiting case? Furthermore, by looking at both sides of the mirror trajectory, can we understand, or at least reveal some additional features, regarding negative energy flux and negative entropy? 
\end{quote}

We shall now explain the reason we would like to consider only trajectories that have no acceleration horizon (hereinafter, we will simply refer to an acceleration horizon as ``horizon'', unless there is a risk of confusion.) The basic utility of various trajectories examined in the moving mirror model with respect to black hole radiation has, like black hole evaporation itself, also been well-known for over forty years.  But the problem of relating Hawking's global construction calculation (see \cite{Raval:1996vt} for a stochastic route) to the physical mechanism responsible for the particle-energy creation effect, which is safely assumed to involve local curvature, has been mostly elucidated by various methods using the \textit{late-time} Davies-Fulling black hole-moving mirror correspondence.  However, with the advent of the \textit{all-time}, exact, black hole-moving mirror correspondence \cite{Good:2016oey}, the physics of the boundary condition effect is more directly related to the curved space effect than previously supposed.  We therefore hope to timely exploit this insight by extending the flat-space model to one with a boundary condition that does \emph{not} contain an asymptotic horizon.  It is clear from basic causality that the existence of a horizon in the future is not essential to the production of particles and energy at early times, and in this calculation we will explicitly demonstrate the particle production by localization of the spectra at early times. (In fact, the same causality argument works for the case of a black hole, i.e. the event horizon is \emph{not} an essential feature for early-time particle creation. See also \cite{1011.5593}.)

Beyond this, we will demonstrate that \emph{the production process at any time need not have an acceleration horizon in order to reach thermal equilibrium}.  The use of wave packets to obtain spectral resolution of thermal Hawking radiation in such a limiting-case horizonless model where the evaporation process stops at late times, has not been investigated before. It is important to recognize that this model is an extension of the particular moving mirror which has a one-to-one correspondence to the exactly solvable black hole case \cite{Good:2016oey}.  The model presented here is novel, in part, because its limit is that of the exact correspondence, which means that a certain acceleration parameter (see below) is appropriately scaled.  A novel result of the removal of the horizon, for this unique trajectory, is that the finite total emission of energy is simple enough that it can be expressed analytically.  Also important to this solution is the introduction of a new method to find other moving mirror solutions which may otherwise be intractable.  

In this work, we are focused on the mirror model. Although the solutions and their novel features could represent various black hole collapse scenarios,
it is premature to make any strong conclusion here. In addition, even though the moving mirrors that we investigate below have no acceleration horizon, \emph{this does not necessary mean that the corresponding ``black holes'' have no event horizon}\footnote{The possibility that black holes may not have an event horizon has been investigated by many authors, see \cite{Kawai:2013mda} \cite{1510.07157} and the references therein.} (or at least a trapped surface of some sort). After all, a main purpose of moving mirror models is to reproduce particle emission of a black hole --- it is the properties of the produced particles that are in correspondence --- one should be very careful if one wishes to identify the mirror trajectories themselves to geometric properties of black holes such as the event horizons\footnote{While it is true that an accelerated observer in Rindler spacetime sees Unruh temperature that is analogous to the Hawking temperature of a black hole, it does not follow that an acceleration horizon of a mirror always corresponds to a black hole horizon. Furthermore, the temperature in the mirror case, is measured by observers far away from the mirror, not a Rindler-like observer \emph{on} the mirror trajectory.}.

The present work is organized in the following manner.
A short introductory treatment of the moving mirror machinery is given in Section (\ref{sec:background}). It shows how to generalize the exactly solvable ``black mirror'' case to remove the horizon. The subject of interest in Section (\ref{sec:trajectory}) is the explicit trajectory of our new mirror solution.  In Section (\ref{sec:energy}) we solve for the energy flux and the total energy production on both sides of the mirror.  The energy flux is shown to have a spike on the left side and a plateau on the right side.  We also quantify the equilibrium temperature.  In Section (\ref{sec:entropy}) we calculate the entanglement entropy flux and confirm it remains consistent with the energy-entropy relation and unitarity.  In Section (\ref{sec:correlations}) we investigate the stress tensor correlations for the model by solving for the correlation ratio $R_1$ exactly.  This helps confirm thermal equilibrium.  In Section (\ref{sec:particles}) we calculate the particle production and investigate the spectral dynamics.  Here we include consistency checks to verify the total energy produced via the stress tensor, agrees with the total energy produced from summing particle quanta.  We find constant emission of particles emitted to the observer on the right for any arbitrary long period of time.   We conclude with some discussions in Section (\ref{sec:conclusions}). Throughout this work, we use the units $G=c=\hbar=k_B=1$. 

%In Section (\ref{sec:metric}) we only briefly touch on possible generalizations of the tortoise coordinate and metric representation for a corresponding black hole.
 
\section{The Machinery of Moving Mirrors}\label{sec:background}

In Section (\ref{sec:somebackground}), we shall first introduce some basic concepts necessary for understanding our construction of a horizonless solution that generalizes the ``black mirror''.   We then discuss the removal of the acceleration horizon in Section (\ref{sec:removehorizon}).  The solution satisfies three criteria: the presence of a horizonless temperature, an appropriately scaled acceleration parameter, and the termination of evaporation at late times.  

\subsection{Some Conventions of Moving Mirrors} \label{sec:somebackground} 
As the simplest example of the dynamical Casimir effect, the moving mirror model also serves as a way to understand black hole evaporation by imposing an external boundary condition in 1+1 dimensions on the quantum field, rather than an external curved spacetime. Consider then, such a boundary that does not accelerate forever, starting and ending at time-like past infinity $i^-$, and time-like future infinity $i^+$, respectively; possessing  asymptotically zero acceleration in both the far past and far future, and always moving slower than the speed of light.  This fully asymptotically inertial mirror will contain no horizon.  Thus, it will contain no pathological acceleration singularity either.  The plot of such a trajectory is in Figure (\ref{fig:Penrose}). A salient pay-off for horizon-removal is that the mirror system, in addition to being unitary (henceforth, by unitarity we always mean unitarity in the broad sense allowed by Bianchi-Smerlak criteria), produces only a finite amount of total energy, as we will demonstrate in Section (\ref{sec:energy}) and Section (\ref{sec:entropy}).

%The use of the forever accelerating observer in the Unruh effect, or a forever accelerating mirror that results in a null horizon, will require the appropriate left-right construction (for example, \cite{Carlitz:1986nh}) to incorporate complete Cauchy surface information.  \\

%\begin{figure}[ht]
%\begin{center}
%\includegraphics[scale=0.8]{PenroseIAMb.eps}
%\caption{\label{fig:Penrose}In this Penrose diagram, the solid black line indicates the asymptotically inertial trajectory, exhibiting no horizon.  The grey line is the eternally thermal moving mirror. }
%\end{center}
%\end{figure}

%\begin{figure}[ht]
%\begin{center}
%\rotatebox{90}{\includegraphics[scale=0.8]{SpacetimeIAMb.eps}}
%\includegraphics[scale=0.4]{pw.jpg}
%\caption{\label{fig:Spacetime} The asymptotically inertial trajectory with a final coasting speed of half the speed of light displayed in the usual spacetime diagram.  The grey dashed lines represent the light cone, and the black dotted-dashed line shows the asymptote of the mirror trajectory.  The trajectory example here is the same as in the conformal diagram, Figure \ref{fig:Penrose}.  For comparison, the grey line indicates the enternally thermal moving mirror, which contains a horizon coinciding with the light cone. }
%\end{center}
%\end{figure}  
  
The quantum field, $\Psi$, is the massless scalar of the Klein-Gordon equation $\Box \Psi = 0$, whose value is zero, $\Psi|_z =0$, when evaluated at the position of the moving mirror, $z(t)$. The modes, $\phi_{\w'}$ and $\psi_\w$, are on equal footing in the sense that they can both be used to expand the field:
\be \Psi = \int_0^\infty \d\w'\; \left[a_{\w'} \phi_{\w'} + a^\dagger_{\w'}\phi^*_{\w'}\right] = \int_0^\infty \d\w\; \left[b_\w \psi_\w + b^{\dagger}_\w\psi^{*}_\w\right]. \ee
The modes are orthonormal and complete and can be exactly solved in the (1+1)-dimensional case:
\be 
\phi_{\w'} = (4\pi \w')^{-1/2} [e^{-i\w' v } - e^{-i\w' p(u)} ],
\ee
\be
\psi_{\w}  = (4\pi \w)^{-1/2} [e^{-i\w f(v)} - e^{-i\w u} ],
\ee
where the functions $p(u)$ and $f(v)$ are the usual notation for the ray-tracing functions, which are intimately related to the trajectory of the mirror $z(t)$ itself, see \cite{Good:2013lca} and Section (\ref{sec:fourusefulfunctions}). The famous Bogoliubov coefficients appear by expanding one set of modes in terms of the other set of modes, 
\be \label{phichi} \phi_{\w'} = \int_0^\infty \d\w\; \left[\alpha_{\w'\w} \psi_{\w} + \beta_{\w'\w}\psi^*_{\w}\right], \ee
\be \label{chiphi} \psi_{\w} = \int_0^\infty \d\w'\; \left[\alpha^*_{\w'\w} \phi_{\w'} - \beta_{\w'\w}\phi^*_{\w'}\right], \ee
where  
\be \alpha_{\w'\w} = (\phi_{\w'},\psi_{\w}), \qquad \beta_{\w'\w} = -(\phi_{\w'},\psi^*_{\w}), \ee
with the flat space scalar product defined in null coordinates, $(u,v)$, by 
%in spacetime coordinates, 
%\be (\phi_{\w'} , \eta_{\w} ) = i \int_{-\infty}^{\infty} dx\; \phi_{\w'}^* \stackrel{\leftrightarrow}{\partial_t}\eta_{\w}, \ee
%in null coordinates,
\be (\phi_{\w'} , \psi_{\w} ) \equiv i \int_{-\infty}^{\infty} \d u\; \phi_{\w'}^* \stackrel{\leftrightarrow}{\partial_u} \psi_{\w} + i \int_{-\infty}^{\infty} \d v\; \phi_{\w'}^* \stackrel{\leftrightarrow}{\partial_v} \psi_{\w}. \ee
The Bogoliubov coefficients $\alpha_{\w'\w}$ and $\beta_{\w'\w}$ also give the operators $a_{\w'}$ and $a_{\w'}^\dagger$ in terms of the operators $b_{\w}$ and $b^{\dagger}_{\w}$, while the orthonormality of the modes hold according to the usual convention, see \cite{Good:2015nja} for more detail. 
%\be \label{aforBog} a_{\w'} =  \int d\w \left[\alpha^*_{\w'\w} b_{\w} - \beta^{*}_{\w'\w} b^{\dagger}_{\w}\right], \ee
%\be \label{bforBog} b_{\w} =  \int d\w' \left[\alpha_{\w'\w} a_{\w'} + \beta^{*}_{\w'\w} a^{\dagger}_{\w'}\right]. \ee
%In null coordinates, for a mirror that has no horizon, the field modes, $\phi_{\w'}$ and $\eta_{\w}$ are normalized on the past and future surfaces respectively,
%\be  (\phi_{\w},\phi_{\w'}) = i \int_{-\infty}^{\infty} dv\; \phi_{\w'}^* \stackrel{\leftrightarrow}{\partial_v} \phi_{\w} = \delta(\w-\w') \ee
%\be  (\eta_{\w},\eta_{\w'}) = i \int_{-\infty}^{\infty} du\; \eta_{\w'}^* \stackrel{\leftrightarrow}{\partial_u} \eta_{\w} = \delta(\w-\w') \ee
%
%The orthonormality of the modes are expressed via,
%\be \label{phinormal}(\phi_{\w},\phi_{\w'}) = -(\phi^*_{\w},\phi^*_{\w'}) = \delta(\w-\w') ,\; (\phi_{\w},\phi^*_{\w'}) = 0, \ee
%\be \label{chinormal}(\psi_{\w},\psi_{\w'}) = -(\psi^*_{\w},\psi^*_{\w'}) = \delta(\w-\w') ,\; (\psi_{\w},\psi^*_{\w'}) = 0. \ee

\subsubsection{The Four Functions of Mirror Physics}\label{sec:fourusefulfunctions}
There are four functions, $v_s(u)$, $u_s(v)$, $x_s(t)$, $t_s(x)$, which are useful for doing global calculations involving the aforementioned field modes.  The first two are the ray-tracing functions (expressed in null coordinates), where $u_s(u) \equiv p(u)$ and $v_s(v)\equiv f(v)$, and the last two are the associated spacetime coordinate functions.  The inverses are expressed like so: 
\be u_s(v) = v_s^{-1}(u),\quad x_s(t) = t_s^{-1}(x).\ee
We shall collectively call all four of them, ``shock wave functions'' or ``shock functions'' for short, after the collapse of the null shell shock wavefront description to form a black hole.  There are many other auxiliary functions, such as $t_s(v)$, $v_s(t)$, $t_s(u)$, $u_s(t)$. However, the original four functions of coordinates $v$, $u$, $t$, and $x$ will prove efficient at calculating observables.  The information about how the field modes become red-shifted due to external conditions is fully contained in these four functions. The relationships between them are demonstrated as follows. 

First consider the usual null coordinates on Minkowski spacetime $ u \equiv t-x$ and $ v \equiv t+x$, and their analogous auxiliary functions as functions of time, 
\be u_s(t) = t- x_s(t), \quad v_s(t) = t+ x_s(t). \label{coordshock} \ee
These contain the shock function $x_s(t)$, which is the trajectory of the mirror.  The inverses of Eqs.~(\ref{coordshock}), contain the shock functions, $u_s(v)$ and $v_s(u)$,
\be t_s(u) = \frac{1}{2}(v_s(u) + u), \quad t_s(v) = \frac{1}{2}(u_s(v) + v). \ee
Functional inverses should be obvious from the notation.  Useful auxiliary inverses are: $t_s(u) = u_s^{-1}(t)$, and $t_s(v) = v_s^{-1}(t)$. The total energy emitted, the energy flux, and the beta Bogoliubov coefficients have expressions that are conveniently written in terms of the four shock functions.
%: $u_s(v)$, $v_s(u)$, $x_s(t)$, $t_s(x)$.
    %In the case of a remnant,  information still comes out -- i.e. the final remnant only contains maybe a few bits of information, or even no information at all.  

\subsection{How to Remove a Horizon}\label{sec:removehorizon}

There is an easy way to remove the horizon, (recall that $c=1$),
\be \lim_{t\rightarrow +\infty}| \dot{\mathfrak{z}}(t)| = 1,\ee
 from a future asymptotically null moving mirror trajectory, $\mathfrak{z}(t)$.  The idea is to modify this so that the new trajectory, $z(t)$, has
\be \lim_{t\rightarrow +\infty}| \dot{z}(t)| = \xi, \ee
where $0<\xi < 1$ is the future asymptotically drifting speed.  This can be achieved by writing the horizonless trajectory, $z(t)$, in terms of the trajectory with a horizon (henceforth `horizon trajectory'), $\mathfrak{z}(t)$, 
\be z(t) =  \xi  \mathfrak{z}(t). \ee
This works if 
\be \dot{z}(t) = \xi \dot{\mathfrak{z}}(t). \ee
Taking this approach helps answer whether the particle spectra can (1) reach equilibrium for an extended period of time, and (2) proceed to shut off.  The mirror does not strictly have a null horizon, yet as we will see, it can still achieve a ``thermal plateau'' (i.e. the emission is virtually thermal for some arbitrary finite amount of time).  This approach also ensures (3) the correct scale for the acceleration parameter $\kappa$ (not to be confused with the physical acceleration, see below).  A correct scale is critical for the red-shifting of the modes to correspond to the exactly solvable black hole case \cite{Good:2016oey, paper1, paper2, Good:2016bsq} in the limit $\xi \rightarrow 1$.  This automatically extends the mirror in the black hole-moving mirror correspondence by promoting it to a more physical footing where the total evaporation energy is finite and unitarity is preserved.  

While we have found the mirror solution that meets these strict requirements, a possible black hole counterpart calculation is beyond the scope of this work.  In the model we are about to present, we do not claim that it actually corresponds to any realistic evaporating black hole spacetime.  For the present work we only seek a simple mirror model in which the three conditions presented above are met, so that we may study the energy and energy flux, the entropy, the correlations, and the particle spectra, together in the absence of a horizon.  It may or may not have an exactly tractable black hole correspondence. In a subsequent work this will be investigated, but as we have emphasized in the Introduction, even if it has such a black hole correspondence, the absence of horizon in the mirror model does not necessary entail the absence of any event horizon or trapped surface in the black hole geometry.
\emph{It is worth pointing out that this mirror solution is new} --- it is the first explicit demonstration of a unitary solution with a thermal plateau\footnote{Other unitary plateau investigations exist, see Section (\ref{sec:conclusions}) for a discussion of one.} that has limiting red-shifting functions which correspond to the black hole-moving mirror system in \cite{Good:2016oey}.

The information contained in the trajectory equation of motion of the mirror is also contained in the shock functions.  The exactly solvable mirror case in \cite{Good:2016oey} has shock functions:
\bea
 v_{s}(u) &=& v_H - \kp^{-1} W\left[e^{\kp(v_H-u)}\right],\\
 u_{s}(v) &=& v - \kp^{-1} \ln \left[\kp(v_H - v)\right], \label{matchingfunction} \\
 x_{s}(t) &=& v_H -t - (2\kp)^{-1} W \left[ 2 e^{2\kp(v_H-t)} \right], \\
 t_{s}(x) &=& v_H -x -\kp^{-1} e^{ x/2\kp}.
 \eea
 %v_{s}(u) &=& v_H - 4M W\left[e^{\frac{v_H-u}{4M}}\right],\\
 %u_{s}(v) &=& v - 4M \ln \left[\frac{v_H - v}{4M}\right], \label{matchingfunction} \\
 %x_{s}(t) &=& v_H -t - 2M W \left[ 2 e^{\frac{-t + v_H}{2M}} \right], \\
 %t_{s}(x) &=& v_H -x -4 M e^{\frac{x}{2 M}}.
The $W$ is the product log or $W$ Lambert function, which commonly appears in thermal equilibrium contexts, e.g. Wien's law.\footnote{The maximum frequency of the (3+1)-dimensional Planck distribution, $\frac{ V \hbar }{\pi^2 c^3} \frac{\omega^3}{e^{\beta \hbar \omega} -1}$, is $\beta \hbar \omega_{\textrm{max}} = 3 + W\left[-\frac{3}{e^3}\right]$, i.e. the famous displacement law $\beta \hbar \omega_{\textrm{max}} = 2.82144$. }  One way to get these is as follows: Firstly, one has the simple form $u_s(v)$ as it is a simple choice for redshifting ray-tracing $f(v)$ function in the mirror case (or from the spacetime matching solution in the null-shell case). This is given.  Secondly, one takes the inverse to get $v_s(u)$.  While easy, as it turns out, it was unhelpful in obtaining the other shock wave functions. The efficient approach is to notice that $u_s(v)$ has a simpler form than $v_s(u)$, so one uses $u_s(v)$ again to write down $t_s(v)$.  The inverse of this can be calculated.  It is, of course, $v_s(t)$. (Note that if one chooses $v_s(u)$ to write down $t_s(u)$ instead, the inverse is not quite as straight-forward to compute, in fact it is much more complicated.)  So, using $v_s(t)$, one is set to write down $x_s(t) = v_s(t) - t$.  Its inverse is, fortuitously tractable, and gives the above expression for $t_s(x)$.  

We shall interchangeably call the horizon trajectory in the mirror analog case the ``black mirror'' \cite{Good:2016bsq} or ``Omex'' for short \cite{Good:2016oey}.  (``Om''-after the Omega constant, $W(1) = \Omega = 0.567$ where $\Omega e^\Omega = 1$, and `ex' after the exponent in $W$ argument.) The acceleration parameter $\kappa$ in the black mirror case can be identified with the surface gravity in the black hole case, $\kappa = (4M)^{-1}$, for all times. 

% We will use this expression as an algebraic relation so we could also use $M$ in place of $\kappa$ in the mirror case. 
%Lest there be confusion, we emphasize that the mirror has no mass and when $M$ is used it will simply be a free parameter, $M>0$.   

The new moving mirror has the following more complicated shock functions:

\be v_s(u) = \frac{2 \xi}{1+\xi}v_H + \frac{1-\xi}{1+\xi}u- \frac{\xi}{\kp}  W\left[\frac{2 e^{\frac{2\kp(v_H-u)}{1+\xi}}}{1+\xi}\right] ,\ee
\be u_s(v) = -\frac{2 \xi }{1-\xi }v_H + \frac{ 1 + \xi }{1-\xi }  v + \frac{\xi}{\kp} W\left[\frac{2 e^{\frac{2\kp( v_H - v)}{1-\xi}}}{1-\xi}\right], \label{matchingfunction2}\ee
\be x_s(t) = \xi  \left(v_H -t -\frac{W\left[ 2 e^{2\kp(v_H-t)} \right]}{2\kp}\right) \label{traj}, \ee
\be t_s(x) = v_H-\frac{x}{\xi } -\frac{1}{\kp} e^{2\kp x/\xi} .\ee

%\be v_s(u) = \frac{1-\xi}{1+\xi}u-4 M \xi  W\left[\frac{2 e^{\frac{v_H-u}{2 M(1+\xi)}}}{1+\xi}\right] + \frac{2 \xi}{1+\xi}v_H,\ee
%\be u_s(v) = \frac{ 1 + \xi }{1-\xi }  v + 4 M \xi  W\left[\frac{2 e^{\frac{ v_H - v}{2 M (1-\xi)}}}{1-\xi}\right]-\frac{2 \xi }{1-\xi }v_H, \label{matchingfunction2}\ee
%\be x_s(t) = -\xi  \left(2 M W\left[ 2 e^{\frac{v_H-t}{2M}} \right]+t-v_H\right) \label{traj}, \ee
%\be t_s(x) = -4 M e^{\frac{x}{2 M \xi }}+v_H-\frac{x}{\xi } .\ee

While these expressions still depend on the primary parameter $\kp$, the intricacy of these expressions arises from the introduction of a second parameter, $\xi$.  Recall that $v_H$ in the black mirror case is the location of the horizon.  We retain $v_H$ for completeness, but make no mistake: the mirror no longer asymptotes to infinite acceleration at a null horizon, located at $v_H$.  
We shall therefore refer to $v_H$ as a ``residual horizon''.  This mirror begins at rest in the far past, and therefore has no initial asymptotic horizon either.  The absence of horizons generates the finite total energy, akin to the notion that evaporating black holes exhale only a finite energy flux\cite{kodama, hiscock1, hiscock2, kuroda}.    

\section{The Domex Trajectory} \label{sec:trajectory}

The motion of the mirror is given by the trajectory Eq.~(\ref{traj}),

\be \label{trajectory} z(t) =  -\xi  \left(\frac{1}{2\kappa} W\left[ 2 e^{-2\kappa t} \right]+t\right), \ee
where $v_H=0$ for simplicity, and $0<\xi<1$ is the final speed of the mirror as $t\rightarrow\infty$.  The motion is initially asymptotically static, $\lim_{t\rightarrow -\infty} \dot{z}(t) = 0$,
and most notably, the mirror does not approach a future asymptotically static resting state because its future asymptotic speed is
\be \lim_{t\rightarrow +\infty}| \dot{z}(t)| = \xi, \ee
making this trajectory future \emph{asymptotically coasting}.  The future drifting feature of this mirror means it is an exact model for a remnant \cite{Good:2015nja,aharonov,Chen:2014jwq} as described by an early anticipation of such solutions by Wilczek in \cite{Wilczek:1993jn}. 

\begin{figure}[!h]
\centering
\mbox{\subfigure{\includegraphics[width=3.0in]{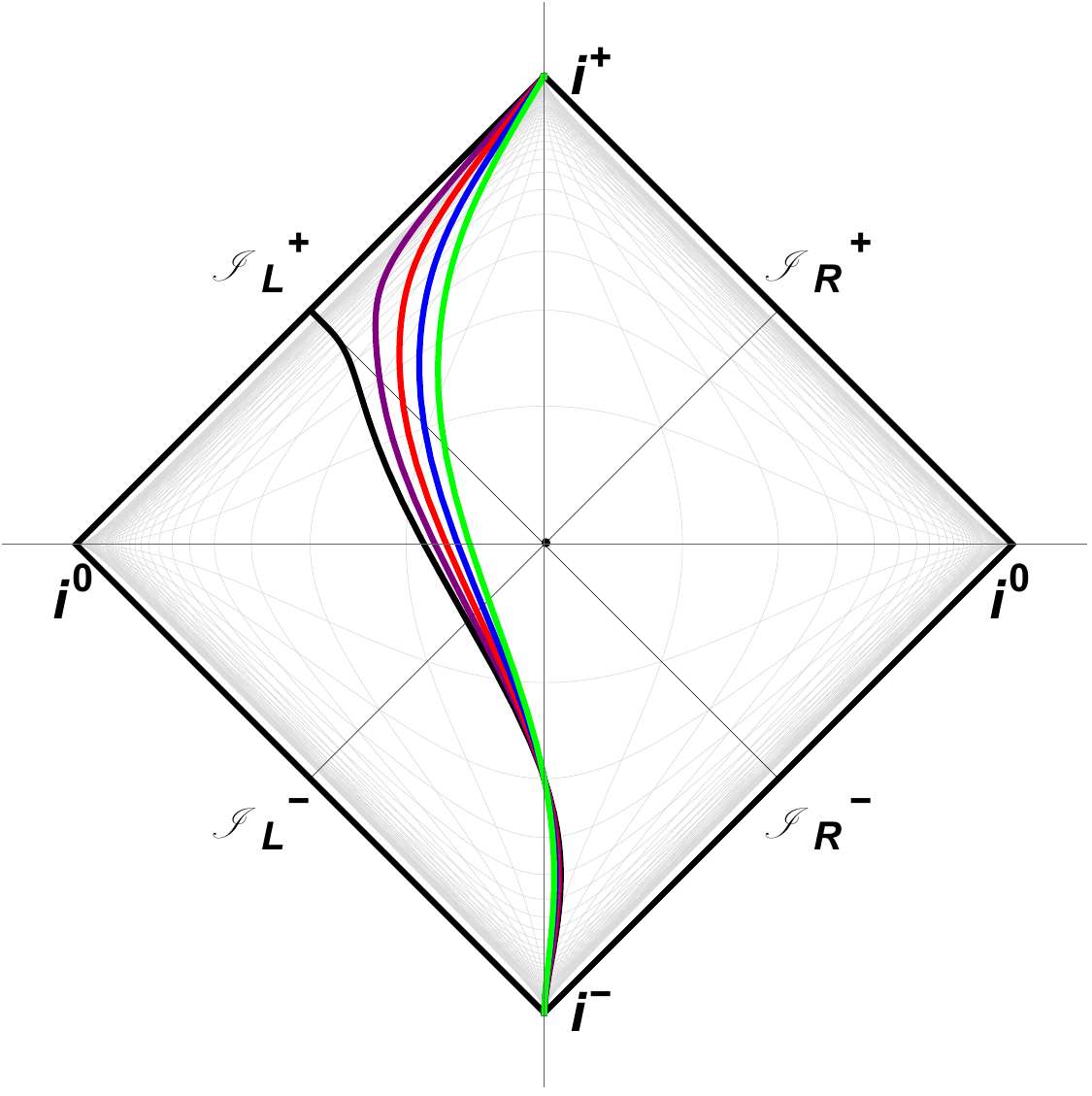}}\quad
\subfigure{\rotatebox{90}{\includegraphics[width=3.0in]{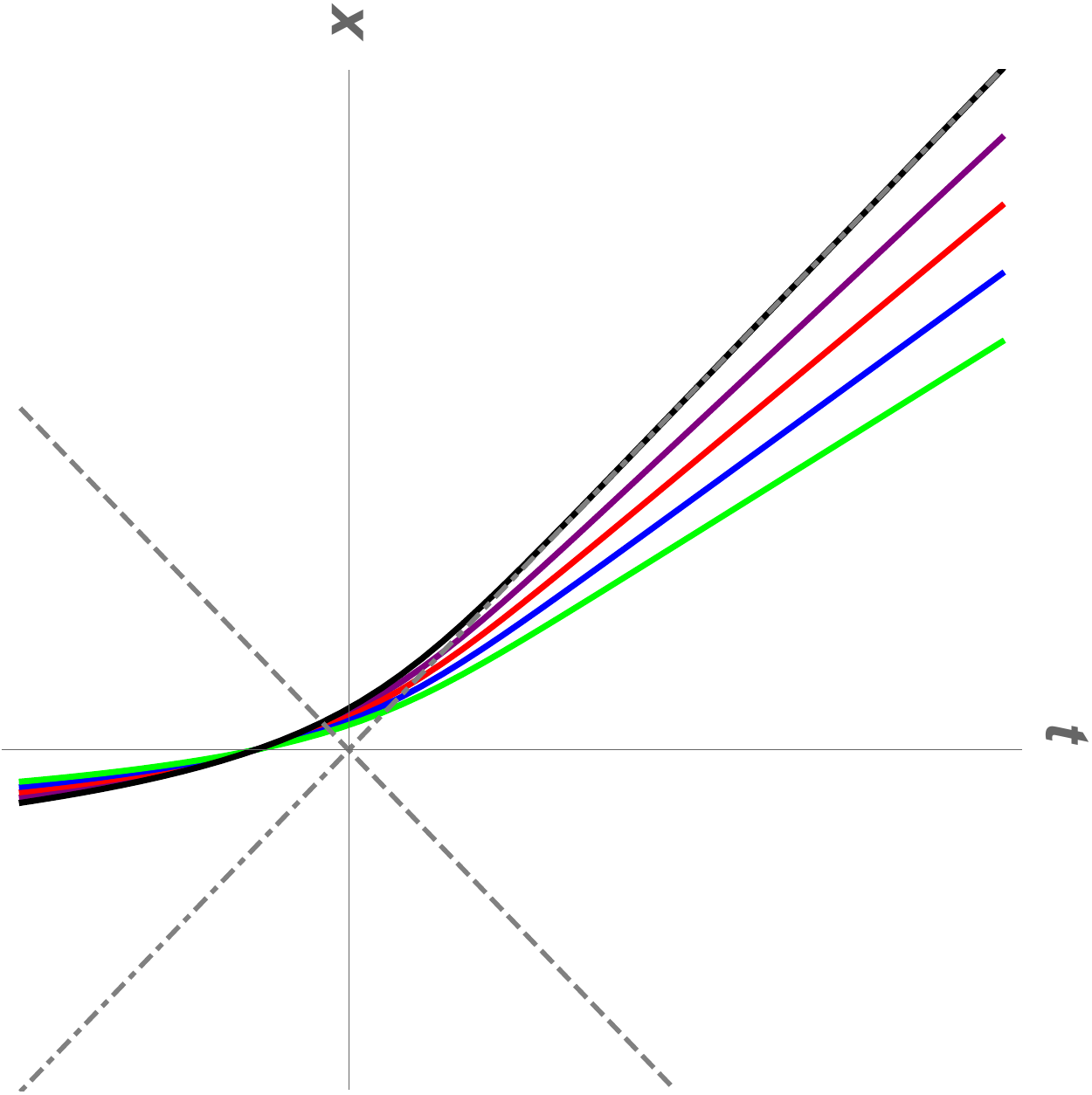} }}}
\caption{\textbf{Left:} In this Penrose diagram, the color curves are ``Domex'', with asymptotically inertial trajectories.  The black curve is ``Omex'',
 a horizon mirror (moving mirror with a horizon) \cite{Good:2016oey}.  The different coasting speeds correspond to $\xi =  0.6, 0.7, 0.8, 0.9$, for green, blue, red, purple, respectively. \textbf{Right:} The asymptotically inertial trajectories (Domex) with the same final coasting speeds displayed in the usual spacetime diagram.  The dashed lines represent the light cone, and the dotted-dashed horizon line is at $v_H=0$.  The trajectory example here is the same as in the conformal diagram.  For comparison, the black line indicates the horizon mirror (Omex) \cite{Good:2016oey}, which contains a horizon coinciding with the light cone. \label{fig:Penrose}} 
\end{figure}

The trajectory Eq.(\ref{trajectory}) is plotted in both the spacetime and Penrose diagrams in Figure (\ref{fig:Penrose}).  The acceleration parameter, $\kp$, is $\kp>0$, and to be clear, it is \emph{not} the acceleration of the mirror, $\alpha(t) \neq \kappa$.  The rectilinear proper acceleration, $\alpha = \gamma^3\ddot{z}$, is time-dependent:
%\begin{widetext}
\be \label{propacc} \alpha(t) = -\frac{2 \kappa  \xi  W\left(2 e^{-2 \kappa  t}\right)}{\left(W\left(2 e^{-2 \kappa  t}\right)+1\right)^3 \left(1-\frac{\xi ^2}{\left(W\left(2 e^{-2 \kappa  t}\right)+1\right)^2}\right)^{3/2}}. \ee
%\end{widetext}
%cannot exceed the maximum in Eqn. \ref{maxacc}.  This maximum sets a coupling limit on $\kp$ and $\xi$ where $\kappa \ll 1$ if $\xi \approx 1$.  
The negative sign on Eq.(\ref{propacc}) gives a mirror whose motion is to the left.  The acceleration has asymptotic behavior such that
\be \lim_{t\rightarrow \pm\infty} \alpha(t) = 0, \ee
making this trajectory asymptotically inertial, despite the drift.  As we shall now show, this solution has several analytically tractable results.  The special physical aspects of this solution will be investigated in the following sections. We shall refer to this horizonless mirror as Drifting-Omex (``Domex'') for short.  

%\cite{Carlitz:1986nh} \cite{Good:2013lca}

%\subsection*{A Specific Example: A Black Hole with Finite Total Evaporation Energy}

%\be u_s(v) = \frac{ 1 + \xi }{1-\xi }  v + 4 M \xi  W\left[\frac{2 e^{\frac{ v_H - v}{2 M (1-\xi)}}}{1-\xi}\right]-\frac{2 \xi }{1-\xi }v_H, \ee
%\be v_s(u) = \frac{1-\xi}{1+\xi}u-4 M \xi  W\left[\frac{2 e^{\frac{v_H-u}{2 M(1+\xi)}}}{1+\xi}\right] + \frac{2 \xi}{1+\xi}v_H\ee
%\be x_s(t) = -\xi  \left(2 M W\left[ 2 e^{\frac{v_H-t}{2M}} \right]+t-v_H\right) \ee
%\be t_s(x) = -4 M e^{\frac{x}{2 M \xi }}+v_H-\frac{x}{\xi } \ee

%These should be compared with the exactly solvable black hole-moving mirror case, as listed previously, where we have shock wave functions:
%\bea
% u_{s}(v) &=& v - 4M \ln \frac{v_H - v}{4M}, \\
% v_{s}(u) &=& v_H - 4M W\left[e^{\frac{v_H-u}{4M}}\right],\\
% x_{s}(t) &=& -t + v_H - 2M W \left[ 2 e^{\frac{v_H-t}{2M}} \right], \\
% t_{s}(x) &=& -x + v_H -4 M e^{\frac{x}{2 M}}.
% \eea

\section{The Energy Production of Domex}\label{sec:energy}

%\subsection*{Four Ways to Calculate the Total Evaporation Energy}

\subsection{The Energy Flux}

 The energy flux of a moving mirror was first derived by Davies and Fulling \cite{Davies:1976hi}.  Expressed in terms of the shock functions, it may be computed via
\bea
F(u) &=& \f{1}{24\pi}\left[\f{3}{2}\left(\f{v_s''}{v_s'}\right)^2-\f{v_s'''}{v_s'}\right],\label{stress}\\
F(v) &=& \f{-1}{24\pi} \left[ \f{3}{2}\left(\f{u_s''}{u_s'}\right)^2 - \f{u_s'''}{u_s'}\right]\f{1}{u_s'^2},\\
F(t) &=& \frac{1}{12\pi}\left[\f{x_s'''(x_s'^2-1)-3x_s'x_s''^2}{(x_s'-1)^4(x_s'+1)^2}\right],\label{stressz}\\
F(x) &=& \frac{1}{12\pi}\left[\f{t_s'''(t_s'^2-1)-3t_s't_s''^2}{(t_s'-1)^4(t_s'+1)^2}\right].
\eea

%\label{stress}
%\be
%F(u) = \f{1}{24\pi}\left[\f{3}{2}\left(\f{p''}{p'}\right)^2-\f{p'''}{p'}\right],
%\ee
%where the primes are derivatives with respect to $u$, and in terms of any mirror trajectory, $z(t)$, is
%\be\label{stressz} F(t) = \frac{1}{12\pi}\left[\f{\dddot{z}(\dot{z}^2-1)-3\dot{z}\ddot{z}^2}{(\dot{z}-1)^4(\dot{z}+1)^2}\right],\ee
%where the dots are time derivatives.
In terms of a $u$-dependent rapidity \cite{Good:2016oey}, $\eta(u) \equiv \tanh^{-1}[\dot{z(t_u)}] =\frac{1}{2}\ln v_s'(u)$, this is 
\be F(u) = \frac{1}{12\pi} \left(\eta'^2 - \eta''\right). \ee
\subsubsection*{Right Side}

\begin{figure}[ht]
\begin{center}
\mbox{\subfigure{\includegraphics[width=3.0in]{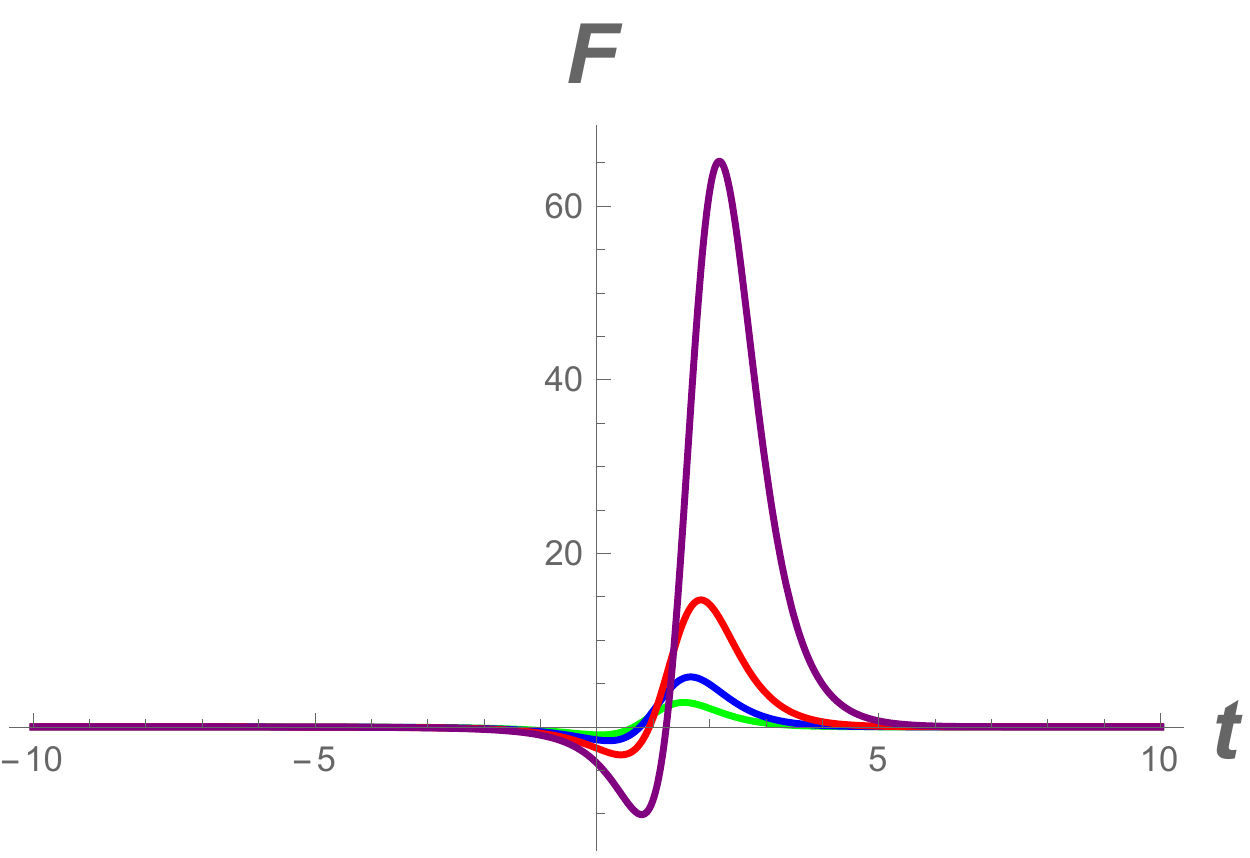}}\quad
\subfigure{\includegraphics[width=3.0in]{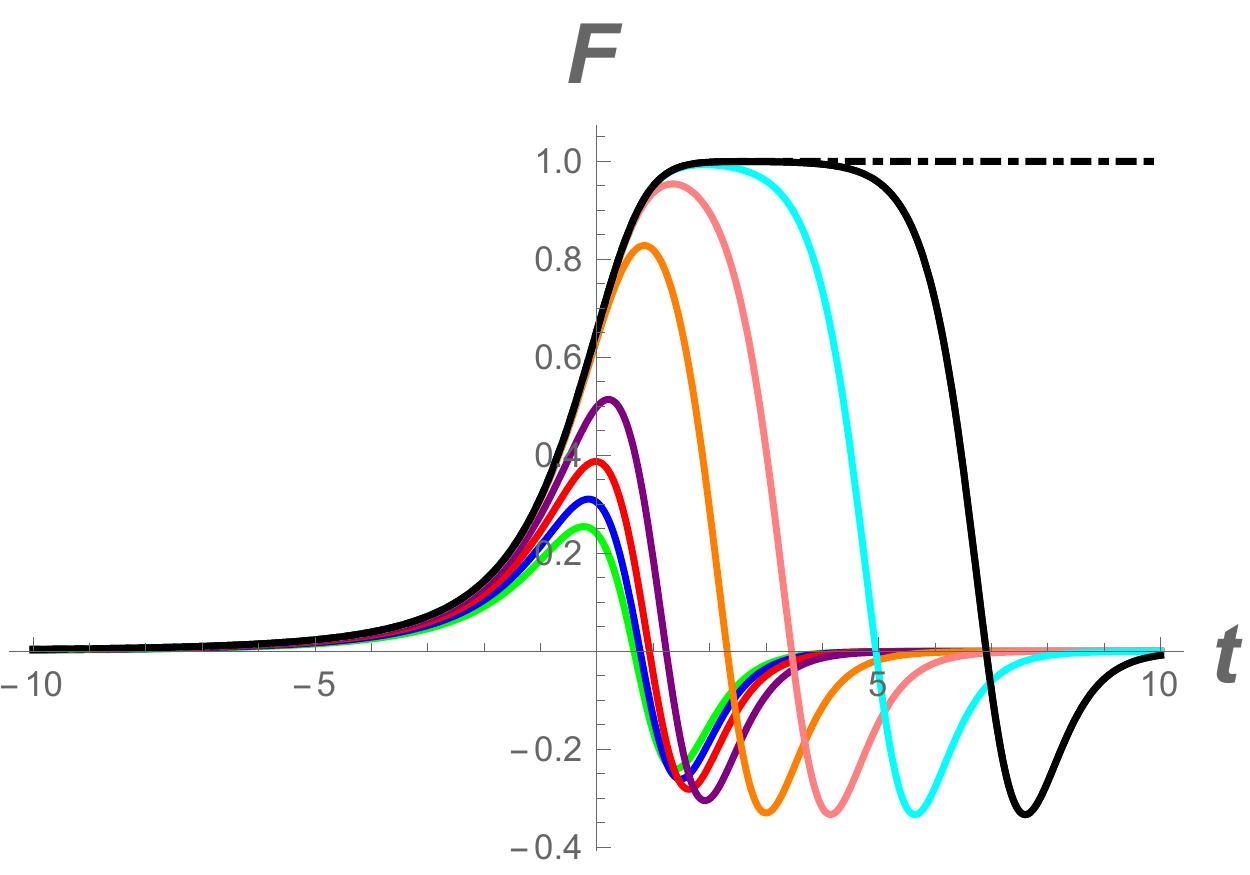} }}
\caption{\label{fig:flux} \textbf{Left:} The left observer sees energy flux that strongly peaks the faster the coasting speed of the mirror: $\xi = 0.6,0.7,0.8,0.9$, colored by green, blue, red, purple, respectively. Notice the initial NEF, and the above-thermal, $F(t)>1$, emission. Here $\kp^2 = 48\pi$.  \textbf{Right:} Successive plots of the energy flux observed by the right observer, $F(t)\equiv\la T_{uu} \ra$, Eq.(\ref{energyflux}), from smallest peak to the largest peak, with varying limiting mirror speeds, $\xi=0.6,0.7,0.8,0.9$. Also included are $\xi=1-0.1^x$ where $x=2,3,4,6$ colored by orange, pink, cyan, black, respectively.  Thermal equilibrium occurs only for a very fast final coasting speed.  The total NEF is qualitatively unchanged at this speed.  The acceleration parameter is set to $\kp^2 = 48\pi$ so that the plateau levels out at $F=1$, the dot-dashed line. }
\end{center}
\end{figure}

The energy flux, emitted to an observer at the right side of the mirror, $\mathscr{I}_R^+$, using the trajectory of Eq.(\ref{trajectory}) in Eq.(\ref{stressz}), is therefore easily calculated:
\be \label{energyflux}F(t) =  \frac{\kappa ^2 \xi  W\left(2 e^{-2 \kappa  t}\right) \left(\xi ^2+2 W\left(2 e^{-2 \kappa  t}\right)^2+W\left(2 e^{-2 \kappa  t}\right)-1\right)}{3 \pi  \left(-\xi +W\left(2 e^{-2 \kappa  t}\right)+1\right)^2 \left(\xi +W\left(2 e^{-2 \kappa  t}\right)+1\right)^4}.\ee
It contains a build-up phase, a thermal plateau, and an end-phase accompanied by negative energy flux (NEF), see Figure (\ref{fig:flux}).  The residual horizon location has been set to $v_H=0$.

A period of thermal emission occurs at extremely high coasting speeds, giving a thermal plateau, which is, in the limit $\xi \rightarrow 1$, located for some time $\Delta t_{TP}$, at  
\be \label{plateau} F(\Delta t_{TP}) \approx F_T \equiv \f{\kp^2}{48 \pi}. \ee 
Interestingly, this is the same as the constant flux produced by the (eternally thermal) Carlitz-Willey trajectory \cite{Carlitz:1986nh}.  The Carlitz-Willey mirror radiates a thermal Planckian distribution of particles for all times, at $F = F_T$.  In our model, this value occurs because in the limit $\xi \rightarrow 1$, this mirror has the same shock functions as the black mirror, which has thermal radiation at late times.  However, now it is apparent that in this model, the evaporation eventually stops, effectively decoupling the late-time approximation from the high-frequency approximation.

Allowing $\xi$ to be nearly the speed of light, (for example, $\xi = 1-0.1^{10}$, with no formal limit), the energy flux emitted to the right observer, has a simple minimal negative value, at some late time, $t_0$, where 
\be F(t_0)^{\text{min}} = -\f{1}{3}F_T, \ee  
which is a fairly significant proportion of the maximum magnitude amplitude of thermal emission.        

\subsubsection*{Left Side}

The energy flux, emitted to an observer at the left of the mirror, $\mathscr{I}_L^+$, using the trajectory of Eq.(\ref{trajectory}) by symmetry reversing the sign on $\xi$, is:
\be \label{energyfluxleft}F(t) = -\frac{\kappa ^2 \xi  W\left(2 e^{-2 \kappa  t}\right) \left(\xi ^2+2 W\left(2 e^{-2 \kappa  t}\right)^2+W\left(2 e^{-2 \kappa  t}\right)-1\right)}{3 \pi  \left(-\xi +W\left(2 e^{-2 \kappa  t}\right)+1\right)^4 \left(\xi +W\left(2 e^{-2 \kappa  t}\right)+1\right)^2}.\ee
The energy flux contains an initial nascent NEF, a rapid reversal to positive energy flux and build-up to a non-thermal positive energy flux peak, and finally a rapid end-phase that falls to zero emission, see Figure (\ref{fig:flux}).
It is now clear that while one side of the mirror is approaching thermal equilibrium emission, the other side is experiencing a single non-thermal, ever-more-narrow burst, demonstrating a characteristic difference between the left and right observers. We investigate the pulse via particle spectra in Section (\ref{sec:particles}).  %\edz{{\color{blue} Who are the corresponding ``left'' and ``right'' observers in the case of black hole? Can one interpret the left side as the black hole interior?}\color{red} yes, seems possible, but unknown, replied in email}

\subsection{Temperature of Domex}
Domex achieves a temperature, $2\pi T = \kappa$, to lowest order in $\epsilon$ where $\xi \equiv 1-\epsilon$, via a ``twice rapid acceleration'' ($\kappa(u) = |p''/p'| = |2\eta'|$) approximation.  The rapid acceleration, $\eta'(u)$, is identically constant, such that $\kappa(u) = \kappa$, for the eternally thermal mirror (Carlitz-Willey).  One finds, 
\be 2\pi T = |2\eta'|= \kappa(1+W(e^{-\kappa u}))^{-2} + \mathcal{O(\epsilon)}. \ee
For large $\kappa u$, so long as, $\kappa u \lll \epsilon^{-1}$, then $W(e^{-\kappa u}) \rightarrow 0$, and to lowest order in $\epsilon$, the rapid acceleration is constant, $2\pi T = |2 \eta'| = \kappa$.
   
From the energy flux production, we can help quantify the equilibrium condition of Domex.  The simplicity of the time-space function, $t_s(x)$, allows for analytic tractability.  Finding where the radiation is most near equilibrium amplitude, $F \approx F_T \equiv \kappa^2/(48\pi)$ is possible.  Using $v_H=0$, and $t_s(x) = -\kp^{-1} e^{2\kp x/\xi} - x/\xi$, one has the flux as a function of space:
\be F(x,\xi,\kp) = \frac{2 \kp^2 \xi  e^{\frac{2\kp x}{ \xi }} \left(8 e^{\frac{4\kp x}{ \xi }}+2 e^{\frac{2\kp x}{\xi }}+\xi ^2-1\right)}{3 \pi \left(-2 e^{\frac{2\kp x}{ \xi }}+\xi -1\right)^2 \left(2 e^{\frac{2\kp x}{ \xi }}+\xi +1\right)^4}. \ee
Maximizing $F(x,\xi,\kp)$ with respect to $x$, gives the spatial location, $x_0$, where the flux is maximum, $F(x_0, \xi, \kp) = F_{\rm{max}}(\xi,\kp)$. Since drift speed is high, then to lowest order in $\epsilon$, ignoring the imaginary component of this spatial locus, the real location is  
\be x_0 = \f{1}{6\kp}  \ln \f{\epsilon}{6} + \mathcal{O}(\epsilon^{1/3}). \ee
The maximum flux, to lowest order in $\epsilon$, is then
\be F(x_0,\xi,\kappa) = F_{\rm{max}} (\xi,\kappa) = \f{\kappa^2}{48\pi}\left[1-3 \sqrt[3]{6} \epsilon^{2/3}+\frac{25}{3}\epsilon + \mathcal{O}(\epsilon^{4/3})\right]. \ee
Following Davies \cite{Davies:1977yv}, Eq. 3.10, or Walker \cite{Walker:1984vj}, Eq. 5.10, %or Reuter-Hill (1989, Eq. 4.37)
we consider the property that the energy flux of a thermal trajectory has
\be F = \int_0^\infty \f{\d\omega}{2\pi} \f{\omega}{e^{\omega/T}-1} = \f{\pi}{12} T^2. \ee
%Domex's radiation is closest to thermal equilibrium when the flux is maximum, $F_m(x_0,\xi,\kappa)$, so that 
Temperature can be expressed as,
\be T(\xi,\kappa) = \sqrt{\f{12}{\pi} F_{\rm{max}}(\xi,\kappa)}, \ee
where we have taken the positive root, $T>0$.  To low order in $\epsilon$, the result is
\be T(\xi,\kappa) = \f{\kappa}{2\pi}\left[1-3\left(\f{3}{4}\right)^{1/3} \epsilon^{2/3} + \f{25}{6} \epsilon + \mathcal{O}(\epsilon^{4/3})\right]. \ee
The lowest order dependence on drift speed scales as $\sim(1-\xi)^{2/3}$, indicating, e.g. that speeds of $\xi = 1-0.1^9$ give a millionth part deviation from equilibrium temperature.  To ensure Domex is very near equilibrium for an extended period of time, we will use speeds far faster while investigating the time dependence of particle production in Section (\ref{sec:particles}).

\subsection{Total Energy Produced by Domex}

It proves possible to calculate the finite total emitted energy analytically. In terms of the shock wave functions, the total energy to the right side of the mirror, is computed via,
\bea
E &=& \int_{-\infty}^{\infty} F(u) \d u,\\
E &=& \int_{-\infty}^{\infty} F(v) u_s' \d v,\\
E &=& \int_{-\infty}^{\infty} F(t) (1-x_s')\d t, \label{fluxtime}\\
E &=& \int_{+\infty}^{-\infty} F(x) (t_s' - 1)\d x. 
\eea
or after integration by parts, where the boundary term is ignored due to asymptotic inertial character,
\bea
E &=& \f{1}{48\pi}\int_{-\infty}^{\infty}\left(\f{v_s''}{v_s'}\right)^2 \d u,\\
E &=& \f{1}{48\pi}\int_{-\infty}^{\infty}\f{u_s''^2}{u_s'^3}  \d v,\\
E &=& \f{1}{12\pi}\int_{-\infty}^{\infty}\f{x_s''^2}{(1+x_s')^2(1-x_s')^3} \;\d t,\\
E &=& \f{1}{12\pi}\int_{-\infty}^{\infty}\f{t_s''^2}{(1+t_s')^2(1-t_s')^3} \;\d x.
\eea
where $v_s\equiv v_s(u)$, $u_s\equiv u_s(v)$, $x_s \equiv x_s(t)$ and $t_s \equiv t_s(x)$.  The primes always mean derivatives with respect to the respective function variable.  For the mirror trajectory here with finite energy production we shall use the $x_s(t)$ integral over $\d t$ and confirm it with quanta summing of particles in Section (\ref{sec:particles}), where the total emitted energy is $E = \int_0^{\infty}\d\w \; \w \int_0^{\infty}\d\w' \;|\beta_{\w\w'}|^2 $.

Note that by ``total'', we mean that it is the total amount of energy that \emph{only} the observer on the right side detects.  The mirror emits energy on both sides to two separate observers: left and right.  To find the energy emitted to the left, by symmetry, one can simply reverse the motion and compute the energy on the right side again.    
%To account for both sides of the mirror, one must add up the total energy on both sides.  
%More explicitly, this gives consistency between Eq.(\ref{energyflux}) and Eq.(\ref{particlebeta}):
%\be \label{quantasum} \int_0^\infty \omega \left[\int_0^\infty |\beta_{\omega\omega'}|^2 d\omega'\right] d\omega = \int_{-\infty}^{\infty} \la T_{uu}(t) \ra (1-\dot{z})dt. \ee 
%which is of course the total energy, Eq.(\ref{totalenergy}). 
%The total energy on a single side of the mirror, may also be found by using the energy flux of the previous section, $F(t)\equiv \la T_{uu}(t) \ra$,
%\be E = \int_{-\infty}^{\infty} F(t) (1-\dot{z})\d t. \ee 
%Expressed in terms of $u$, it can be calculated simply as \cite{ Good:2013lca} ,
%\be E = \int_{-\infty}^{\infty} F(u) \d u. \ee 
%In our model, this can be done analytically.  

\subsubsection{Right Side}
%Consider pairing energy, $\w$, with each particle emitted.
The total energy radiated to $\mathscr{I}_R^+$ is therefore:
\be \label{rightsideenergy} E_R = \frac{\kappa  (3- \xi) \tanh ^{-1}(\xi )}{48 \pi  \xi ^2}-\frac{\kappa  (3+ 2\xi)}{48 \pi  (\xi^2 + \xi)}.\ee
%\frac{\kappa  (3- \xi) \tanh ^{-1}(\xi )}{48 \pi  \xi ^2}-\frac{\kappa  (2 \xi +3)}{48 \pi  \xi  (\xi +1)}.\ee%= -\frac{\kappa }{48\pi} \frac{\left(\xi  (2 \xi +3)+(\xi -3) (\xi +1) \tanh ^{-1}(\xi )\right)}{\xi ^2 (\xi +1)} \ee
Domex does not result in the emission of infinite energy to the usual observer at $\mathscr{I}_R^+$.  Note that the solution here is monotonic for increasing coasting speed and never negative for $0<\xi<1$.  Here, the $\lim_{\xi\rightarrow 0} E_R = 0$, and the $\lim_{\xi\rightarrow 1} E_R = +\infty$. 

\subsubsection{Left Side}
For an observer to the left at $\mathscr{I}_L^+$ the total energy emitted is found by simply substituting, $\xi\rightarrow -\xi$, into $E_R$, 
\be E_L = \frac{\kappa  (3+ \xi) \tanh ^{-1}(-\xi )}{48 \pi  \xi ^2}-\frac{\kappa  (3- 2\xi)}{48 \pi  (\xi^2 - \xi)}. \ee
%\frac{\kappa  \left(4 \xi ^2-6 \xi +(\xi -1) (\xi +3) \log \left(\frac{2}{\xi +1}-1\right)\right)}{96 \pi  (\xi -1) \xi ^2}.\ee%= -\frac{\kappa }{48\pi} \frac{\left(\xi  (2 \xi +3)+(\xi -3) (\xi +1) \tanh ^{-1}(\xi )\right)}{\xi ^2 (\xi +1)} \ee
Again, the energy is finite as long as the speed is less than the speed of light.  The expression, $E_L(\xi)$ is a monotonic function of $\xi$.

\subsubsection{Both Sides} 
\begin{figure}[ht]
\begin{center}
\mbox{\subfigure{\includegraphics[width=3.0in]{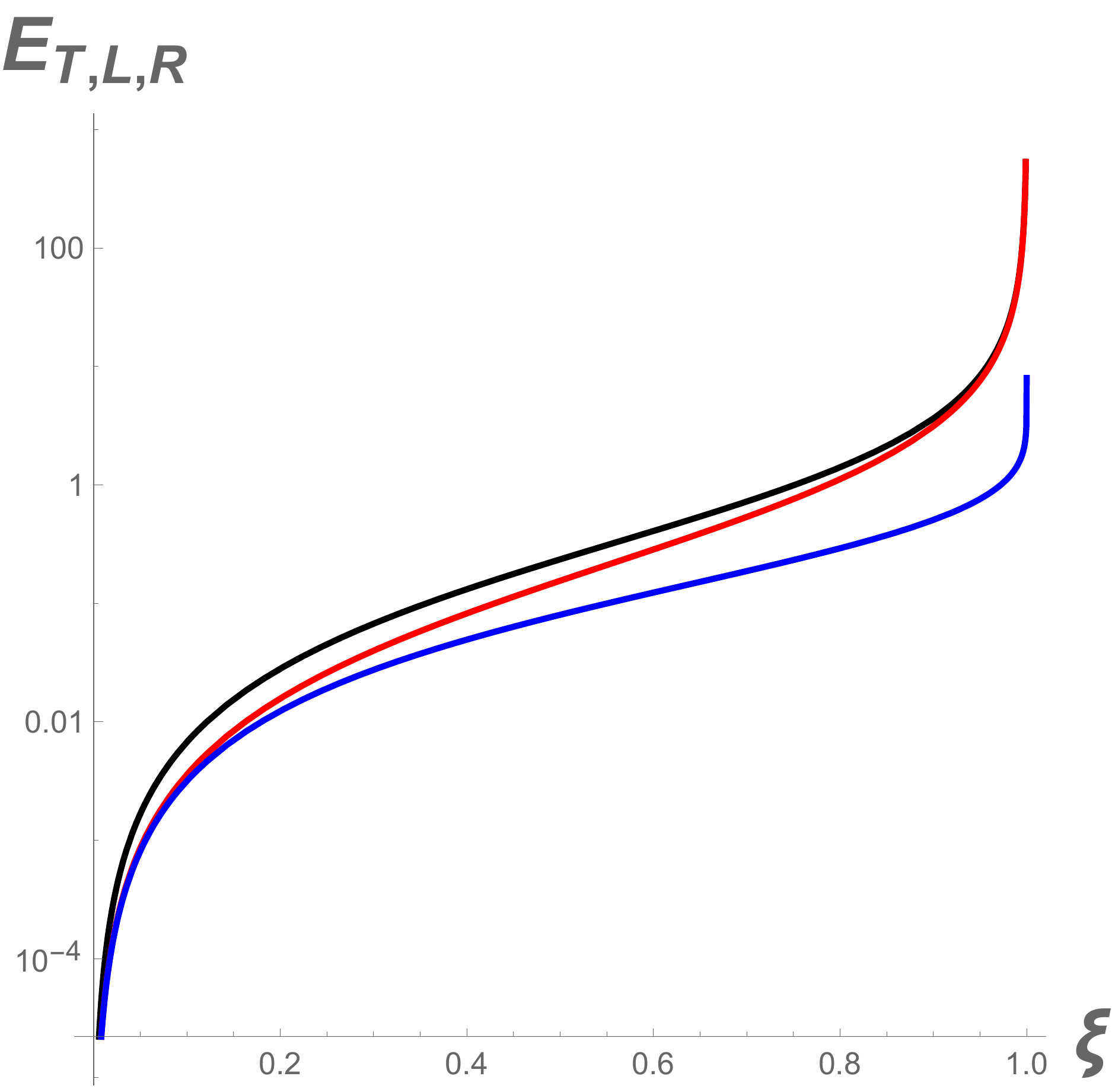}}\quad
\subfigure{\includegraphics[width=3.0in]{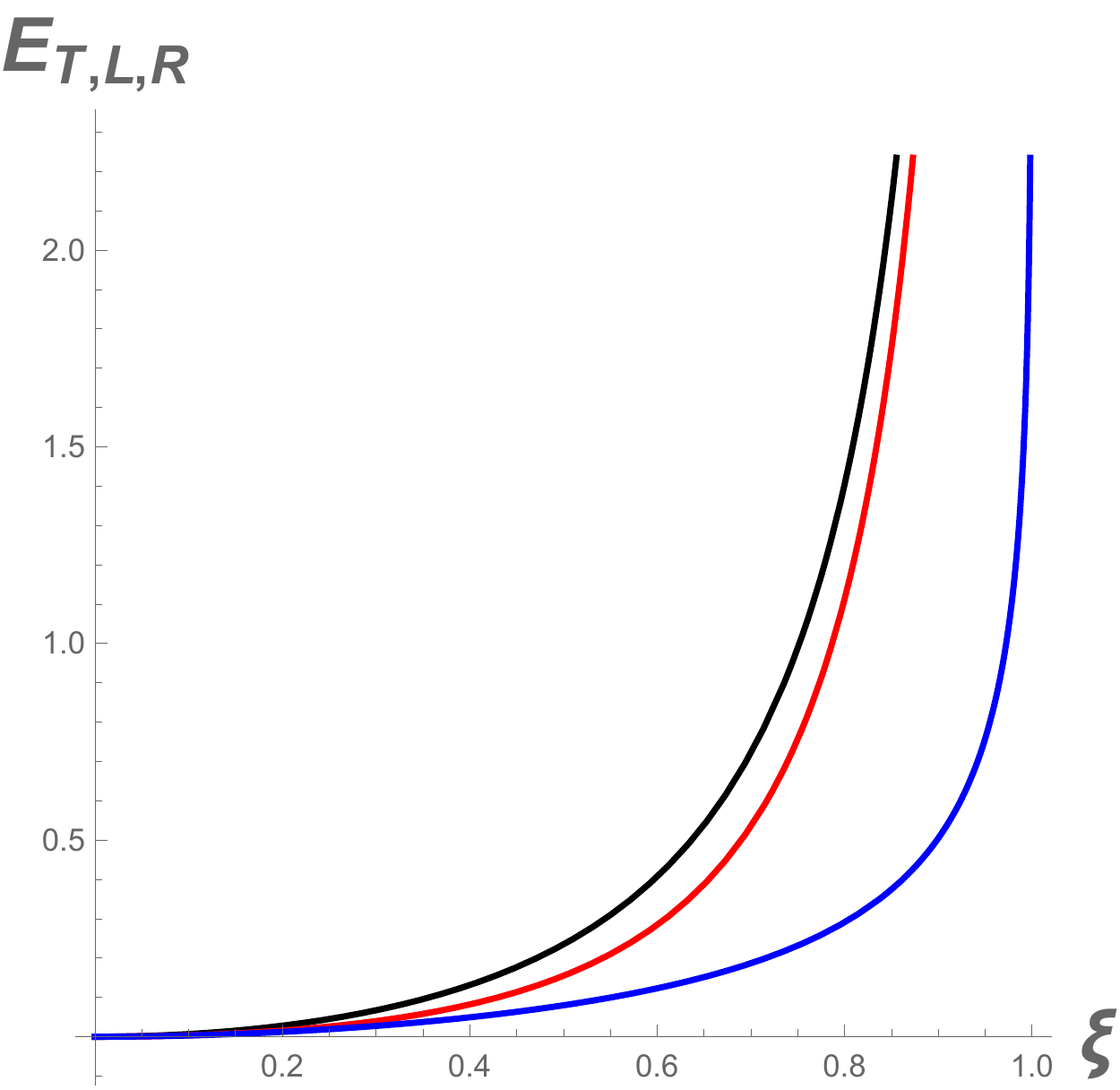} }}
\caption{\label{fig:totalenergy} \textbf{Left:}  Energies plotted in a semi-log plot.  The total energy emitted is the black line, $E_T(\xi) = \frac{\kp}{24\pi}(\frac{1}{1-\xi^2}-\frac{\tanh^{-1}\xi}{\xi})$, with $\kp = 24\pi$ in a log plot. The energy diverges as the final coasting speed approaches the speed of light. The red line is the $E_L$ and the blue line is $E_R$. \textbf{Right:} The same energies, $E_T$, $E_L$, and $E_R$, plotted to scale, $\kp = 24 \pi$. } 

\end{center}
\end{figure}

For the high coasting speeds we are interested in, the energy emitted to the left is \emph{always much} greater than the energy emitted to the right, $E_L\gg E_R$.  For small values of $\xi$ one finds
\be \frac{E_L}{E_R} = 1 + \frac{6}{5} \xi + \mathcal{O}(\xi^2), \ee
indicating that $E_L > E_R$.  As it turns out, $E_L > E_R$ for all values of the final drift speed, $0< \xi < 1$.    
%\edz{{\color{red} KY demonstrated this with Resolve and ForAll command in MMA}}
The total energy emitted to both observers is $ E_T = E_L + E_R$:
\be E_T = \frac{\kappa}{24\pi}\left[\cosh ^2(\eta )-\eta  \coth (\eta )\right], \ee
where $\eta = \tanh^{-1} \xi$, is the final rapidity. See Figure (\ref{fig:totalenergy}) to see a graph of the combined total emitted energy from both sides of the mirror.  Notice the divergent behavior as the coasting speed approaches the speed of light.  The energy increases monotonically as a function of the coasting speed. 

\section{The Entropy of Domex}\label{sec:entropy} 
 
\begin{figure}[ht]
\begin{center}
\mbox{\subfigure{\includegraphics[width=3.0in]{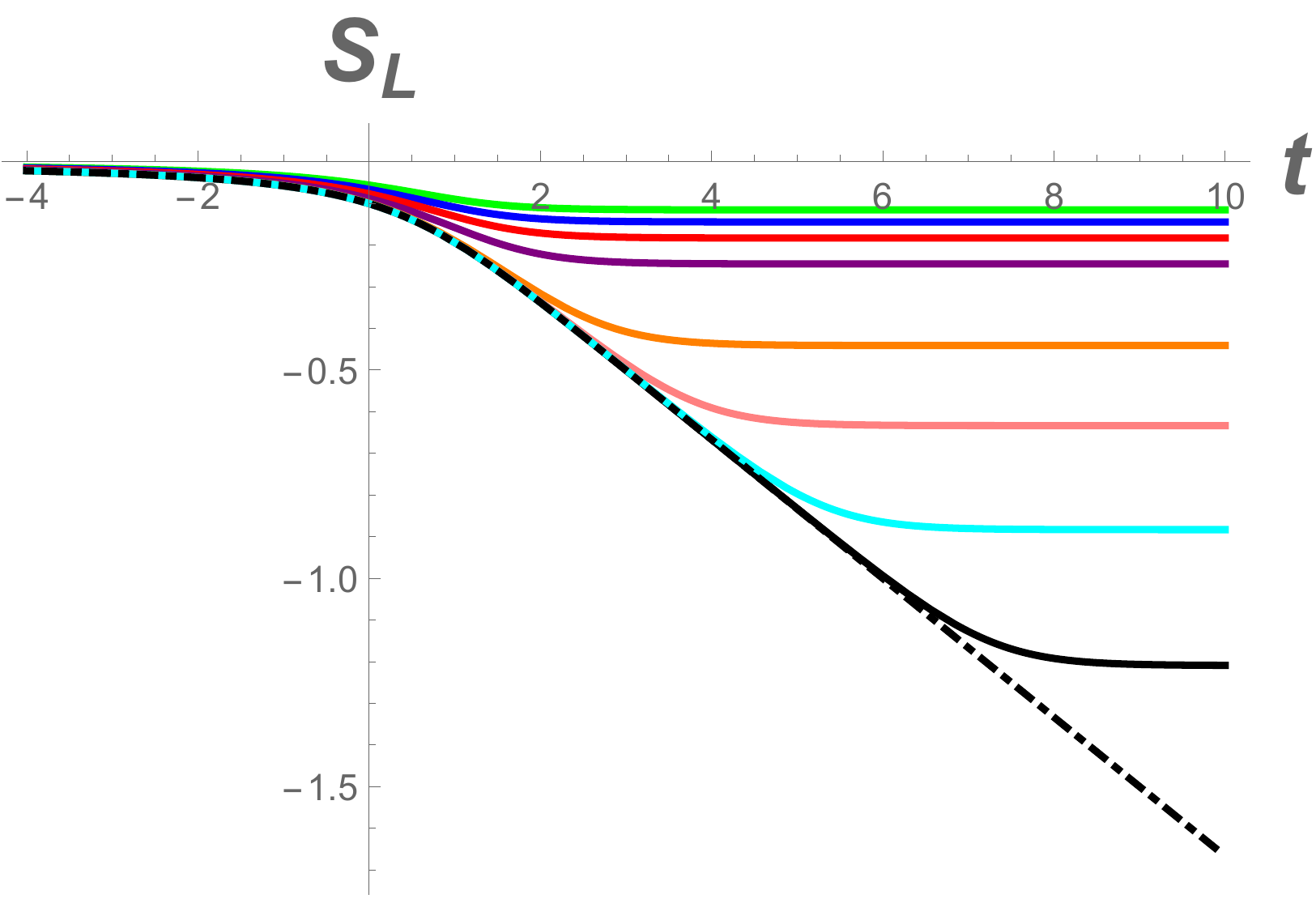}}\quad
\subfigure{\includegraphics[width=3.0in]{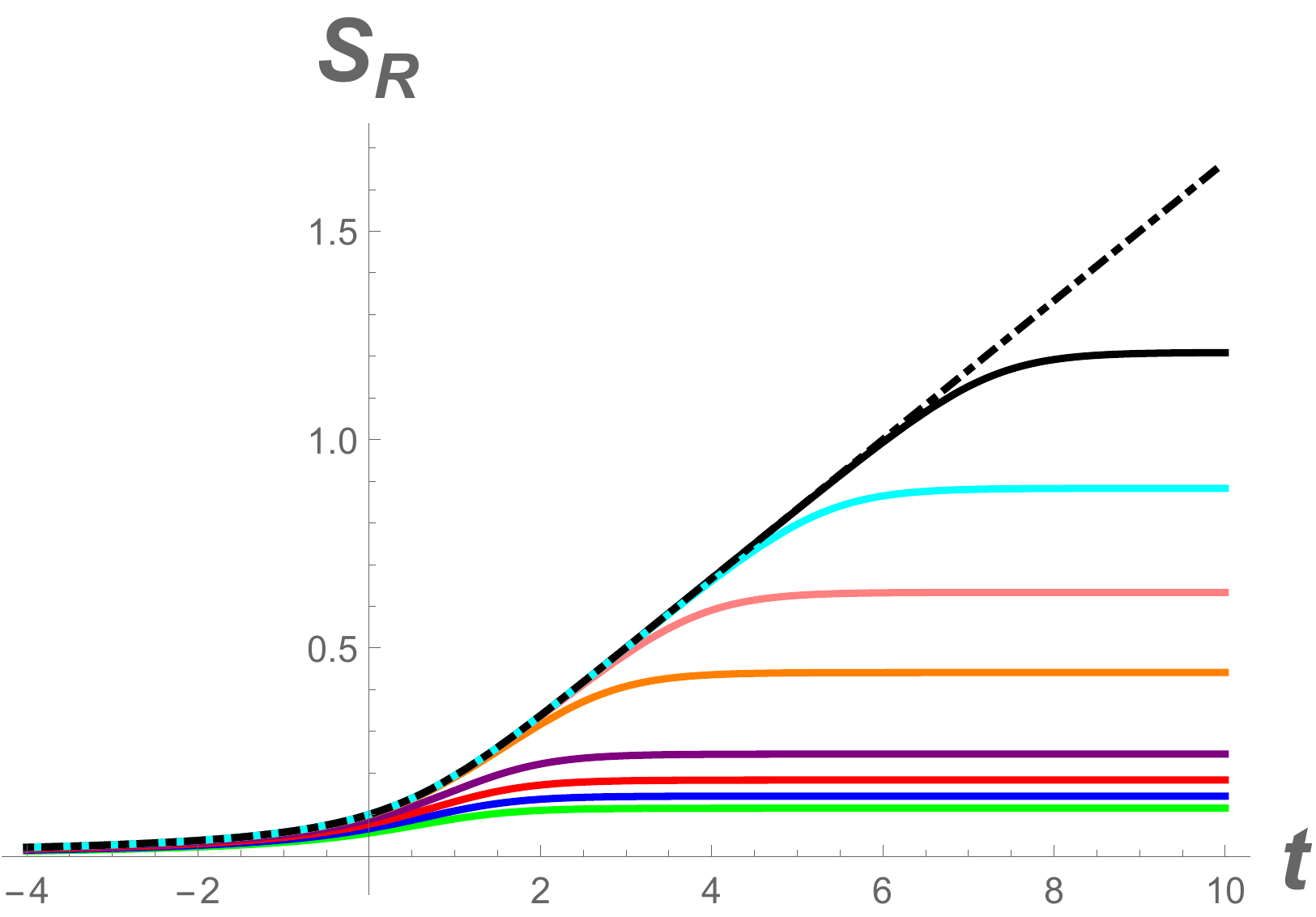} }}
\caption{\label{fig:entropy} The von Neumann entanglement entropy, for the left and right sides of the mirror respectively, as a function of time, with $\xi$ values and colors as the previous Figure (\ref{fig:flux}) but here $\kappa = 1$.  The entropy approaches a constant value, underscoring the final asymptotic coasting speed of the mirror and the eventual end of particle emission, with no turn over of the Page curve.  All left moving field modes become right moving field modes, and vice versa, preserving unitarity. } 
\end{center}
\end{figure}

The von Neumann entanglement entropy in the unitary moving mirror case, can be found from Bianchi-Smerlak's formula\footnote{Here the central charge for a conformal field theory, $c$, has been set to unity without loss of generality.} \cite{Bianchi:2014qua} as a function of $v_s(u) \equiv p(u)$,
\be S(u) = -\frac{1}{12} \ln p'(u). \ee 
In terms of the mirror trajectory (see Good-Ong \cite{Good:2015nja}), this is
\be \label{entropy} S(t) = -\frac{1}{6} \tanh^{-1} [\dot{z}(t)] = -\frac{1}{6}\eta(t), \ee
where $z(t)$ is the trajectory motion of the moving mirror, the dot represents the time derivative and $\eta(t) \equiv \tanh^{-1}[\dot{z}(t)]$ is the time-dependent rapidity.  It is simple to see that the faster the mirror moves, the greater the entropy.   Unitarity in this sense strictly means that the entropy must achieve a constant value in the far past and far future.  The mirror we consider has a non-zero asymptotic entropy, underscoring that eternally thermal radiation is not the end state.  Since evolution from a possible initially pure state, to a final mixed state does not occur, radiation of energy flux stops and it is possible to reestablish a possible initial pure state, preserving unitarity.  
%The energy flux may also be expressed at the mirror in terms of the entropy as a function of time:
%\be 2 \pi F(t) = e^{-12 S(t)} \cosh^2 [6 S(t)] ( 6 \dot{S}(t)^2 \tanh[6 S(t)] + \ddot{S}(t)), \ee
This mirror's trajectory is found to be consistent with the Bianchi-Smerlak \cite{Bianchi:2014qua} entropy-energy relationship 
\be 2\pi F(u) = 6 S'(u)^2 + S''(u), \ee
where the entropy for Domex, expressed as a function of time, $t$, for both the right and left sides,
\be \label{entropyLR} S_L^R(t) = \pm \frac{1}{6} \tanh ^{-1}\left(\frac{\xi }{W\left(2 e^{2\kappa(v_H-t)}\right)+1}\right), \ee
approaches a constant value,
\be \lim_{t\rightarrow \infty} S_L^R(t) = \frac{1}{6} \tanh ^{-1}(\xi ) = \pm \frac{\eta}{6}, \ee
as the energy flux approaches zero.  Notice the independence of $\kappa$ or the position of the horizon $v_H$ for the final entropy value.  Here $\eta$ is the final coasting rapidity ($\xi$ is the final coasting speed). These results are in dramatic contrast to Omex \cite{Good:2016oey} which has infinite total energy and a divergent entanglement entropy in the far future, $t\rightarrow +\infty$. A plot of the entropy (on both sides of Domex), Eq.~(\ref{entropyLR}), is in Figure (\ref{fig:entropy}).  

For Domex, the entanglement entropy has no turn-over of the Page curve \cite{page1, page1b, page2} as it is monotonically increasingly approaching some constant value.  Although outside the scope of this work, it is worth mentioning that the radiation in an analogous black hole context would never get purified.  The preservation of unitarity is maintained only in the sense that the pure state remains pure taking into account both the exterior and interior \cite{aharonov} of the black hole remnant. There are unresolved challenges associated with this type of scenario stemming from the known infinite production problem, stability, and energy conservation. Detailed discussions are provided in \cite{Chen:2014jwq}. Note however, that Domex, as a possible candidate model for a remnant, is not in conflict with Wilczek red-shifting \cite{Wilczek:1993jn} and non-monotonic mass loss of Bianchi-Smerlak \cite{Bianchi:2014vea}, as the former occurs even when Domex is once again inertial and in the later, the non-monotonic mass loss occurs as long as the entropy approaches a constant (not necessarily zero). 

We note that the entanglement entropy of the radiation on the left side of the mirror is negative. This is not unprecedented in the literature, however its interpretation is somewhat unclear. Bianchi, De Lorenzo, and Smerlak interpreted a negative entanglement entropy as the result of the radiation being less correlated than the vacuum \cite{1409.0144}. On the other hand, Holzhey, Larsen, and Wilczek were of the opinion that negative renormalized entropy is the result of the radiation having \emph{more} correlation than the vacuum \cite{9403108v1}. We leave open the interpretation of our results, but it is interesting to note that for any fixed $\xi$, the positive entanglement entropy on the right side of the mirror \emph{exactly} cancels the negative entanglement entropy on the left side. Therefore, the entanglement entropy of the radiation on the \emph{entire slice} of any constant $t$ is zero.
\newline

\section{The Correlations in the Radiation}\label{sec:correlations} 

\subsection{Correlation Functions}

An under-appreciated lesson stressed by Ford and Roman \cite{Ford:2004ba}, is that there is a great deal more happening in the accelerating mirror geometry than is revealed by the expectation value of the stress-energy tensor alone. There are subtle increases or reductions in correlations between the flux along rays even where the expectation value vanishes.  The stress-energy tensor correlation function is of interest in our situation because it reveals information about the energy flux that demonstrates the thermal character of the radiation above and beyond that of the thermal plateau of the stress-energy tensor during equilibrium.  The shock functions for the moving mirror are needed to compute the correlation functions for the stress-energy tensor. %\edz{\color{red} ... difference between correlation functions and entropy. Entropy measures the degree of entanglement in the radiation.... correlation? ..entropy approaches a constant.., yet correlation increases? \color{blue}{The main difference in these relationships is $S\sim\ln p'$ and $R_1 \sim f(p,p')$.  B.S.'s relationship for entropy is mostly unrelated to the variance of the stress tensor energy flux. Its more directly related to the shock function derivative $p'$, a more fundamental object.   } } 
It was previously shown that the ray-tracing function $p(u)$ is useful for delta-function pulse piece-wise mirror trajectories \cite{Ford:2004ba}.  In this section, we extend this work to continuous trajectories and compute the correlations with an emphasis on the equilibrium period of Domex.  The energy fluxes emitted by any moving mirror can be positive and negative, but they are only average values.  The fluctuations around this average value are generally expected because the quantum state is not an eigenstate of the stress-energy tensor operator. 

The correlation function for the stress-energy tensor is
\be C_{\mu\nu \mu'\nu'} = \la T_{\mu\nu}(y)T_{\mu'\nu'}(y')\ra - \la T_{\mu\nu}(y)\ra \la T_{\mu'\nu'}(y')\ra ,\ee
where the spacetime points are indicated by $y=(u,v)$ and $y' = (u',v')$.  The correlation functions between two right moving rays, two left moving rays, and right and left moving rays are, respectively:
\be C_{RR}(u,u') = \la T_{uu}(u)T_{uu}(u')\ra - \la T_{uu}(u)\ra \la T_{uu}(u')\ra, \ee
\be C_{LL}(v,v') = \la T_{vv}(v)T_{vv}(v')\ra - \la T_{vv}(v)\ra \la T_{vv}(v')\ra,\ee
\be C_{LR}(v,u') = \la T_{vv}(v)T_{uu}(u')\ra - \la T_{vv}(v)\ra \la T_{uu}(u')\ra. \ee
Solved in terms of the ray tracing function, $p(u)$, the results are \cite{Ford:2004ba}
\be C_{RR}(u,u') = \f{[p'(u')p'(u)]^2}{8\pi^2[p(u')-p(u)]^4}, \ee
\be C_{LL}(v,v') = \f{1}{8\pi^2[v'-v]^4}, \ee
\be C_{LR}(v,u') = \f{[p'(u')]^2}{8\pi^2[p(u')-v]^4}, \ee
where $p'(u) = \d p(u)/\d u$ and $p'(u')=\d p(u')/\d u'$.  

The above expressions deal only with correlations of distinct rays.  These expressions simplify, as would be expected, in vacuum or with a static mirror present.  For a static mirror we have the condition, $v = p(u) = u$, and 
\be C_{RR}(u,u') = C_{\text{vac} \oplus \text{static}}(u,u') = \f{1}{8\pi^2[u'-u]^4}, \ee
\be C_{LL}(v,v') = C_{\text{vac} \oplus \text{static}}(v,v') = \f{1}{8\pi^2[v'-v]^4}, \ee
\be C_{LR}(v,u') = C_{\text{static}}(v,u') =  \f{1}{8\pi^2[u'-v]^4}. \ee
In vacuum $C_{LR}(v,u') = 0$ because there can only be correlations with left and right moving fluxes with a mirror present.  The correlation limits for $C_{RR}$ and $C_{LL}$ hold for either vacuum or a static mirror, hence the xor, $\oplus$, in the subscript.  The ratios 
\be \label{R1} R_1 \equiv \f{C_{RR}(u,u')}{ C_{\text{vac} \oplus \text{static}}(u,u') },\ee
and
\be \label{R2} R_2 \equiv \f{C_{LR}(v,u')}{ C_{\text{static}}(v,u')} \ee
can tell us about enhancement and suppression of correlations.  For $R_i>1$ one interprets enhancement, for $R_i<1$ there is suppression. 
\subsection{Correlation Solutions}

We focus on one ratio only in order to help confirm thermal equilibrium to the right observer.  This correlation ratio, $R_1$, involving $C_{RR}$, associates two right moving rays.  These rays come off the mirror heading to the right observer at $\mathscr{I}_R^+$.  
The $R_1$ solutions for the ratios for the three mirrors of interest,
\begin{enumerate}
	\item[(1)] Thermal Mirror (Carlitz-Willey)
	\item[(2)] Black Mirror (Omex)
	\item[(3)] Horizonless Mirror (Domex)
\end{enumerate}
are included here for completeness.  They are respectively,
\be R_1^{\textrm{Thermal}} = \frac{\kappa ^4 (u-u')^4 e^{2 \kappa  (u+u')}}{\left(e^{\kappa  u}-e^{\kappa  u'}\right)^4}, \ee
which, illustrates thermal correlations at all times.  However, for the black mirror, we have
\be R_1^{\textrm{Omex}} = \frac{\kappa ^4 (u-u')^4 W\left(e^{-\kappa u}\right)^2 W\left(e^{-\kappa u'}\right)^2}{\left(W\left(e^{-\kappa u}\right)+1\right)^2 \left(W\left(e^{-\kappa u'}\right)+1\right)^2 \left(W\left(e^{-\kappa u}\right)-W\left(e^{-\kappa u'}\right)\right)^4}, \ee
which, illustrates thermal correlations at late times.  Finally, we write down the Domex's ratio, 
\be R_1^{\textrm{Domex}} = \frac{\kappa ^4 (u-u')^4 \left(-\xi +(\xi +1) W_u+1\right)^2 \left(-\xi +(\xi +1) W_{u'}+1\right)^2}{\left(W_u+1\right)^2 \left(W_{u'}+1\right)^2 \left(\kappa  (\xi -1) (u-u')+\xi  (\xi +1) W_u-\xi  (\xi +1) W_{u'}\right)^4},\ee
where $W_u \equiv W\left(\frac{2 e^{-\frac{2 \kappa  u}{\xi +1}}}{\xi +1}\right)$, and  $W_{u'} \equiv W\left(\frac{2 e^{-\frac{2 \kappa  u'}{\xi +1}}}{\xi +1}\right)$.  

\begin{figure}[ht]
\begin{center}
\mbox{\subfigure{\includegraphics[width=3.0in]{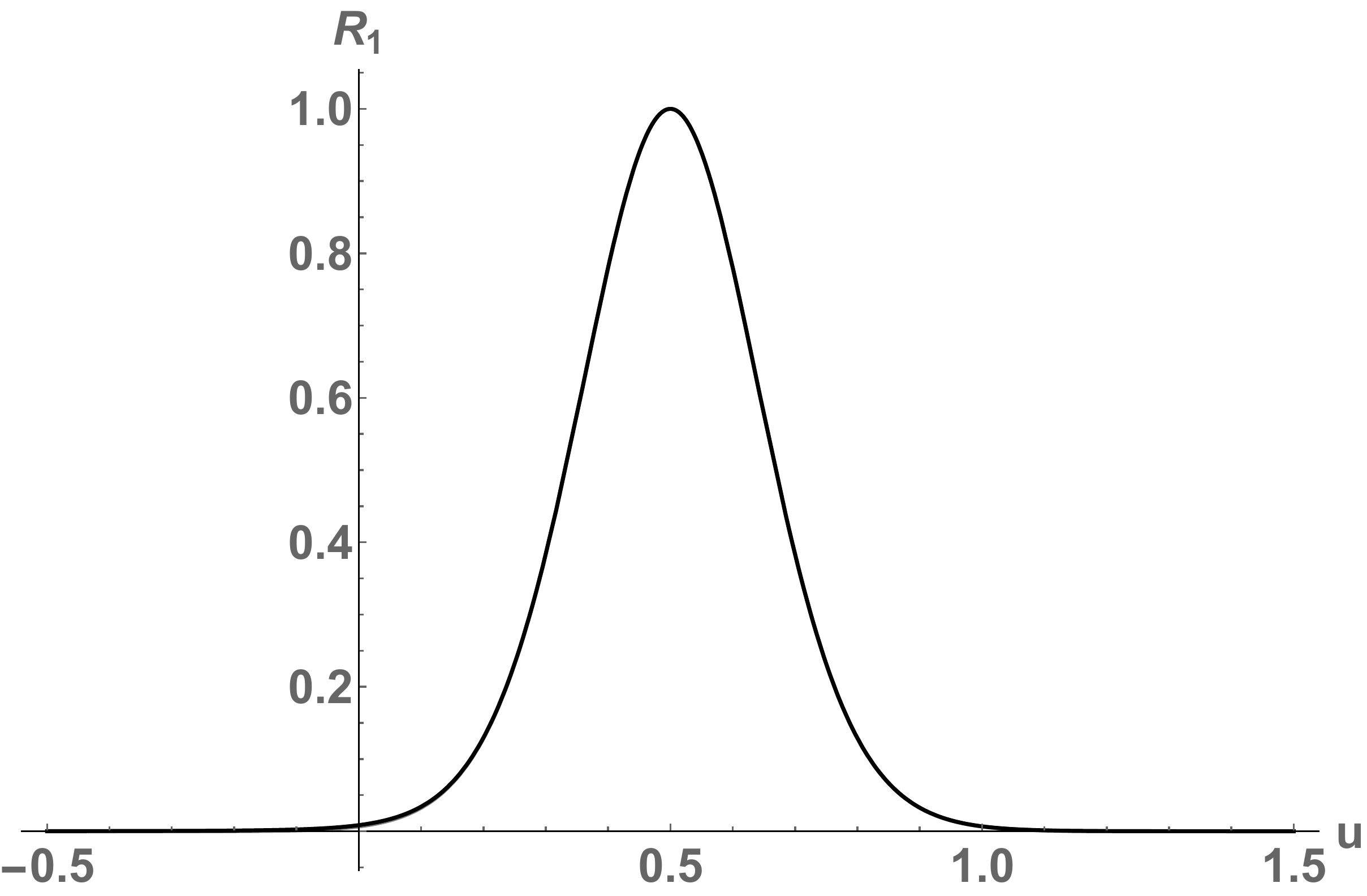}}\quad
\subfigure{\includegraphics[width=3.0in]{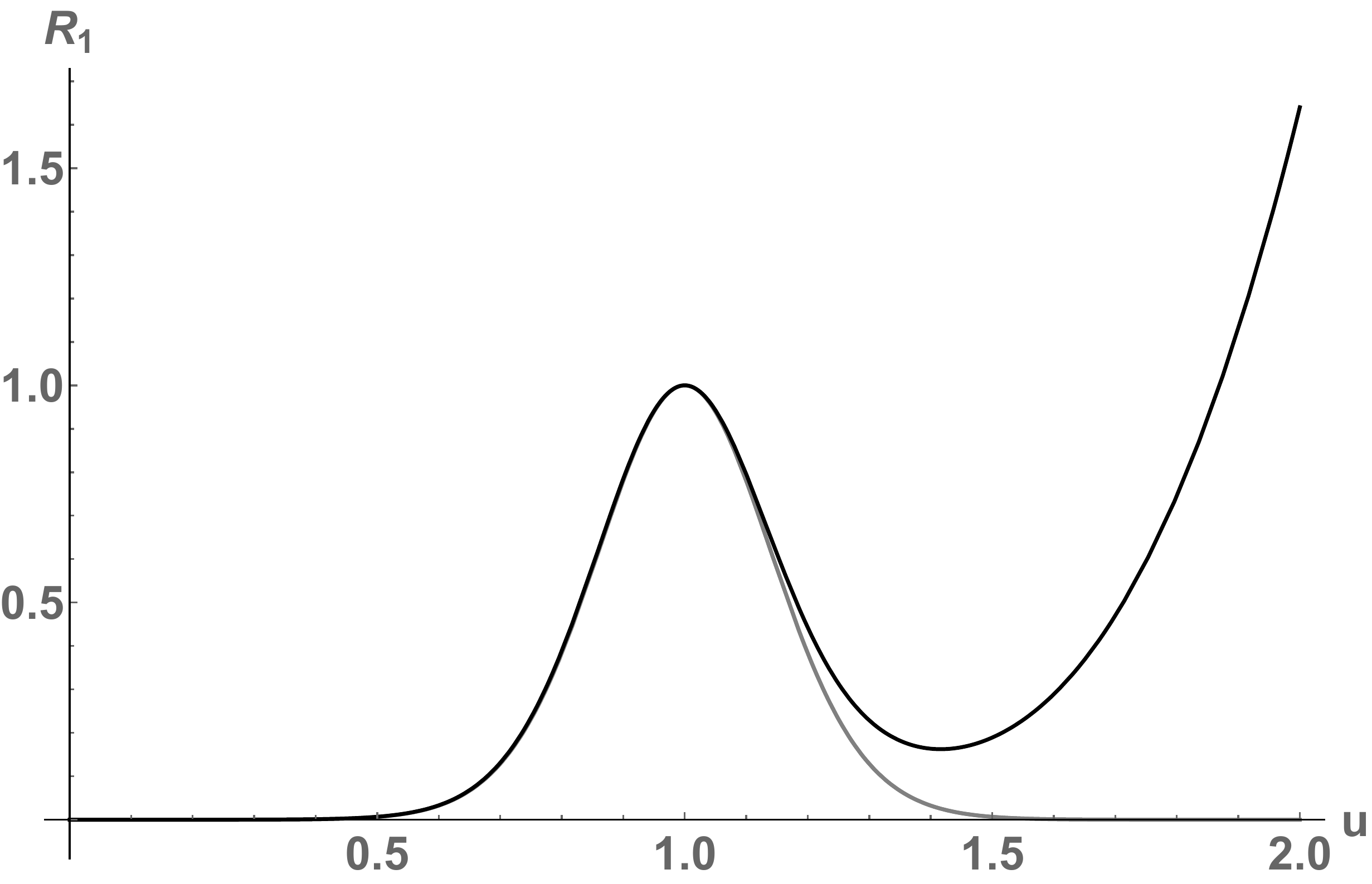} }}
\caption{\label{fig:R1thermal} \textbf{Left:} The ratio $R_1$ for all three mirrors: Carlitz-Willey, Omex and Domex ($\xi = 1-10^{-7}$), where the right moving ray $u' = 0.5$ (chosen for simplicity) and $\kappa^2 = 48\pi$ (chosen to normalize a thermal energy flux of $F_T = \kappa^2/48\pi = 1$).  Notice all three correlations stack up closely on each other, indicating no difference in the correlations from thermal emission between all three mirrors.  With this $\kappa$ scale the maximum flux is near the ray $u'_0 = 0.48 \approx 0.50$ for Domex.  \textbf{Right:} The ratio $R_1$ for all three mirrors where the right moving ray $u' = 1$ and $\kappa^2 = 48\pi$.  The gray line is the thermal mirror and Omex mirror.  The black line is Domex.  Now we are comparing late time correlations, so late in fact, that Domex has correlations that now deviate from thermal equilibrium.  This deviation signals the non-equilibrium completion of evaporation. Notice all three correlations stack up closely on each other only for early values of $u$, indicating no difference in the correlations from thermal emission for these rays.  The ratio becomes enhanced, $R_1>1$, once negative energy radiation is emitted.  } 
\end{center}
\end{figure}

All three mirrors give the same thermal correlations when comparing a ray that occurs in the appropriate equilibrium period. This is not at very late times for Domex.  When one picks a very late time ray, then Omex and Carlitz-Willey are still thermal, but Domex begins to break pattern to abide by the inevitable non-equilibrium completion of emission.  This is to be expected because at very late times the mirror abandons thermal character as the radiation ceases. See Figure (\ref{fig:R1thermal}).

\section{The Particle Production of Domex}\label{sec:particles} 
\subsection{The Beta Bogoliubov Coefficient Integrals}

The distinction between energy flux and particle flux has been well-studied by Walker-Davies\cite{Walker:1982} and Walker\cite{Walker:1984vj}.  The key ingredient is the information of the mirror trajectory equation of motion, which is encapsulated in the shock functions. One needs the Bogoliubov coefficients because the particle emission detected by an observer is
\be \la N_{\omega} \ra \equiv \la 0_{\rm{in}} | N^{\rm{out}}_\omega | 0_{\rm{in}} \ra = \int_0^{\infty} |\beta_{\omega \omega'}|^2 \; \d \omega'. \label{particlecount} \ee
 There are four ways to calculate the beta Bogoliubov coefficient using the shock functions: 

\be
\label{betap}
\beta_{\w\w'} = \f{1}{4\pi\sqrt{\w\w'}}\int_{-\infty}^{\infty} \d u \;
e^{-i\w  u - i\w'v_s(u)} \left(\w' \frac{ \d v_s(u)}{\d u} - \w \right) \;,
\ee

\be
\label{betaf}
\beta_{\w\w'} = \f{-1}{4\pi\sqrt{\w\w'}}\int_{-\infty}^{\infty} \d v \;
e^{-i\w' v -i\w u_s(v)} \left(\w \frac{d u_s(v)}{\d v} -\w' \right) \;,
\ee

\be
\label{betaz}
\beta_{\w\w'} =
\f{1}{4\pi\sqrt{\w\w'}}\int_{-\infty}^{\infty} \d t \;
e^{-i\w_{p}t + i\w_{n} x_s(t)} \left(\w_{p}\frac{ \d x_s(t)}{\d t}-\w_{n}\right) \;,
\ee
\be \beta_{\w\w'} =
\f{-1}{4\pi\sqrt{\w\w'}}\int_{+\infty}^{-\infty} \d x \;
e^{i\w_{n} x -i\w_{p}t_s(x)} \left(\w_{n}\frac{\d t_s(x)}{\d x} - \w_{p} \right) \label{betat}\;.
\ee
Here $\w_p \equiv \w + \w'$ and $\w_{n} \equiv \w-\w'$. The integration bounds also assume all light rays hit the mirror and propagate to future null infinity on the right.  For light rays that do not hit the mirror, one must stop short the integration and add up only to the last null ray as described by the relevant variable.  All of the bounds are written for a mirror which starts at positive spatial infinity and proceeds left to negative spatial infinity in the far future.  The bounds must be appropriately changed for mirrors which have different behaviors and/or horizons.  For Domex we use Eq.~(\ref{betat}) because of the simplicity of $t_s(x)$ function particular to Domex. Notice the negative sign analogous to Eq.~(\ref{betaf}) drops away because the mirror starts at $x \rightarrow +\infty$ at $t \rightarrow -\infty$.  This integral can be used to obtain other trajectories when the other integrals are intractable.  We choose this integral because it will allow investigation of particle production while avoiding integration with the product log, (see \cite{Good:2013lca} for $z(t)$ approach), and therefore insert Domex's trajectory, Eq.~(\ref{trajectory}), into the integral Eq.~(\ref{betat}) identifying, $t_s(x) = t(z)$. 
%\be \label{Bfromt}\beta_{\w\w'} = \f{1}{4\pi\sqrt{\w\w'}}\int_{-\infty}^{\infty} \d z \; e^{i\w_{n} z -i\w_{p}t_s(z)} \left(\w_{n}\frac{\d t_s(z)}{\d x} - \w_{p} \right).
%\ee
%This integral is evaluated analytically with the trajectory of Eq.(\ref{trajectory}).  
The solution is %in terms of an Euler integral of the first kind, i.e. the beta coefficient is a beta function:
%\begin{widetext}
\be \label{particlebeta} \beta_{\w\w'\xi} = -\frac{\xi  \sqrt{\w \w'} (-i (\w+\w')/\kp)^{-\frac{i}{2\kp} ((1+\xi)\w+(1-\xi)\w')} }{2 \pi \kp (\w+\w')}\Gamma \left(\frac{i}{2\kp} ((1+\xi)\w+(1-\xi)\w' )\right) .\ee
%\end{widetext}
 %However they are closely related by the trajectory of the mirror. The information encoded in the trajectory provides the link between the energy flux and the particle production.  It is already known that a mirror can radiate zero, or even negative, energy flux while still producing particles \cite{Walker:1984vj}.
This exact beta solution also contains the thermal plateau, similar to the instantaneous energy flux which closely approaches the thermal line at $F = F_T$.  Both particle and energy flux approach the thermal plateau only at high final speeds making it a salient feature of the radiation.  One should be confident the behavior shows up in both the particle production and energy flux because the packetized particles carry signatures of the instantaneous energy flux emission \cite{Good:2015nja}.  
%This is a stronger evidence than that of the consistency agreement of the bare total energy emission from the summation of quanta energy of Eq.(\ref{quantasum}) alone.  

\begin{figure}[!h]
\centering
\mbox{\subfigure{\includegraphics[width=3.0in]{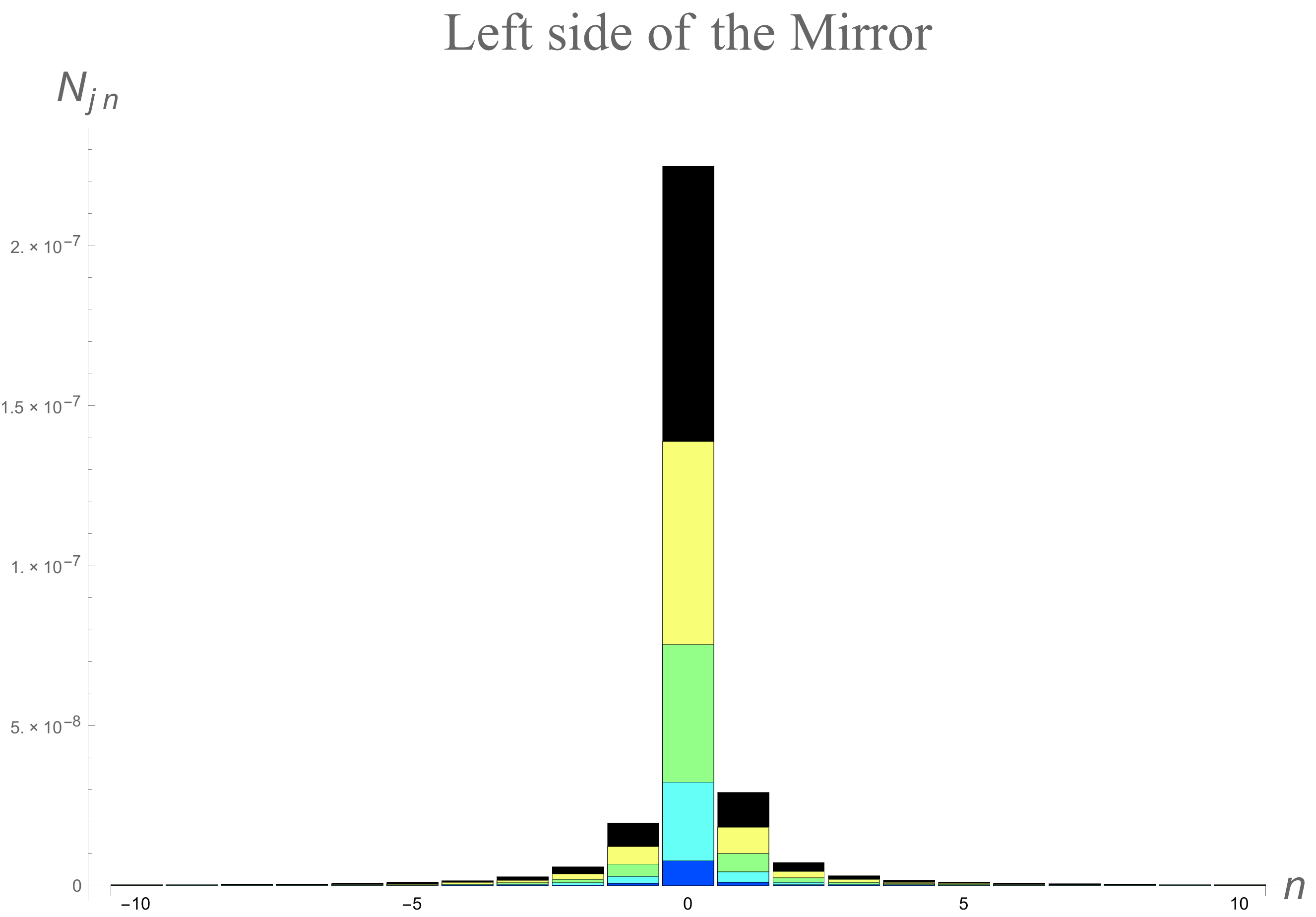}}\quad
\subfigure{\includegraphics[width=3.0in]{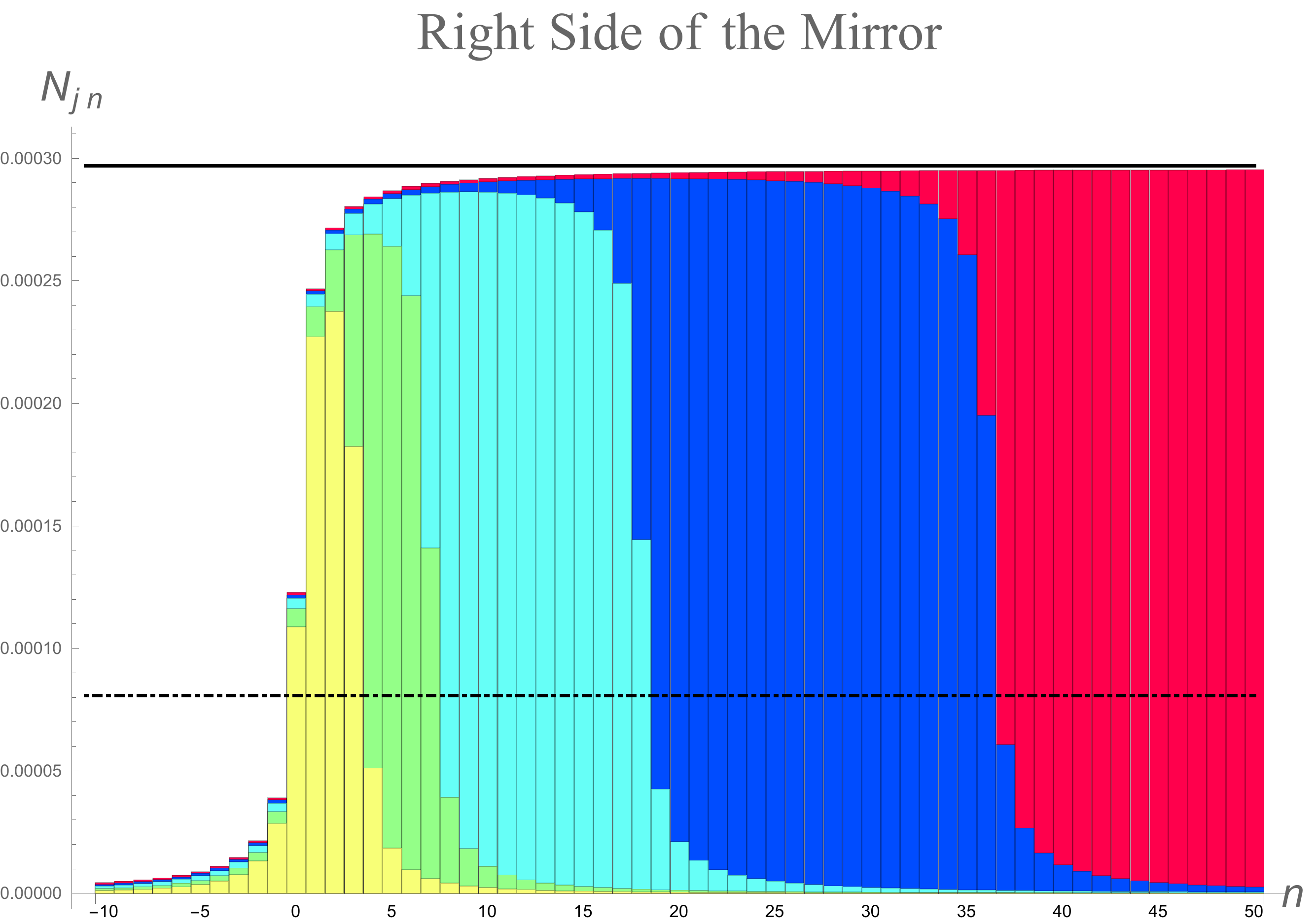} }}
\caption{\textbf{Left:} The left observer sees a non-thermal pulse of particle production.  The final drifting speed of Domex need not be very high to see the growth of the pulse.  Here $\xi = 0.01 a $ where $a = 1,2,3,4,5$.  The acceleration parameter is set to $\kappa = 1$ and the lowest frequency bin $j=1$ is observed. Here $\epsilon = 1$.  The pulse of particles grossly exceeds constant Planckian emission with fast speeds. \textbf{Right:} The right observer sees a plateau of particle production.  The final drifting speed and other parameters are also $\xi = 1-0.1^{x}$, $\kappa=1$, $j=1$, and $\epsilon = 1$, where $x=10,20,50,100$.   Increasing the final coasting speed produces a flatter, more extended in time, thermal plateau. The particle production colored red is Omex, which is thermal at late times ad-infinitum.  The dotted-dashed line is the approximate Planck distribution, $N_j = (e^{\omega_j/T}-1)^{-1}$ where $\omega_j = (j+1/2)\epsilon$. The solid black line is the exact Planck distribution, $N_j = T\epsilon^{-1} \ln \left[ \frac{e^{(j+1)\epsilon/T}-1}{e^{j\epsilon/T}-1}\right] -1$, see e.g. \cite{Good:2016oey}, \cite{Good:2016bsq}. \label{fig:ppfast}} 
\end{figure}

In Figure (\ref{fig:ppfast}), we construct localized beta Bogoliubov coefficients from the global beta Bogoliubov coefficients in Eq. (\ref{particlebeta}),
\be
\beta_{jn\w'} =
\f{1}{\sqrt{\epsilon}}\int_{j\epsilon}^{(j+1)\epsilon}
\d\w \; \left[e^{\frac{2\pi i \w n}{\epsilon}} \beta_{\w\w'\xi} \right]\;.
\label{beta-packet}
\ee
These are the usual orthonormal complete wave packets \cite{Hawking2} which are used to find the time-frequency localized particle count,
\bea  
\la N_{jn}\ra &=& \int_0^\infty \d\w' |\beta_{jn,\w'}|^2   \;,  
\nonumber 
\\
&=& \int_0^\infty \d\w' \int_{j \epsilon}^{(j+1)\epsilon} 
\frac{\d \w_1}{\sqrt{\epsilon}} \int_{j \epsilon}^{(j+1)\epsilon} 
\frac{\d \w_2}{\sqrt{\epsilon}} \left[ e^{\frac{2 \pi i(\w_1- \w_2)n}{\epsilon}}
\beta_{\w_1 \w'} \beta^{*}_{\w_2 \, \w'} \right] \;. 
\label{Njn} 
\eea
%Particles that arrive at $\mathscr{I}^+_R$ can be investigated by obtaining these wave packet coefficients straightforwardly from the 
%coefficients $\beta_{\w\w'}$ of Eq.(\ref{particlebeta}). With the Bogoliubov coefficients for the packets in hand, the average number 
%of particles produced for given values of $n$ and $j$ is determined. 
 Particles arrive at $\mathscr{I}^+_R$ in the range of frequencies $j \epsilon \leqslant \omega \leqslant (j+1) \epsilon$ and in the range of times $ (2\pi n - \pi)/\epsilon \lesssim u \lesssim (2 \pi n + \pi)/\epsilon$.  Details on how to construct these packets in more general situations can be found in \cite{Good:2013lca}.  %There are no signatures of negative energy flux in the particle emission.  Therefore one would not perceive an analog black hole's increase in mass via particle emission like was explored in Good-Ong \cite{Good:2015nja}.  The lack of signatures of NEF in the particle production is not surprising in light of the previous study.

%The horizontal line is the exact solution to particle production with constant flux thermal emission \cite{Good:2013lca}

\subsection{Global Particle Distribution}
The horizonless beta solution Eq.~(\ref{particlebeta}) of Domex has distribution,
\be |\beta_{\w \w' \xi}|^2 = \frac{\omega'}{\pi  \kappa  \left(\omega' + \omega \right)^2}\frac{\xi ^2 \omega }{\left((1+\xi) \omega + (1 -\xi ) \omega '\right)}\frac{1}{ \left(e^{\frac{\pi}{\kappa}  \left((1+\xi) \omega + (1-\xi) \omega'\right)}-1\right)},\ee
while the horizon beta Bogoliubov coefficients of Omex have particle count per mode-squared, %(see Good-Anderson-Evans \cite{Good:2013lca}),
\be |\beta_{\w \w'}|^2 = \frac{\w'}{2\pi \kappa (\w+\w')^2}\frac{1}{e^{2 \pi \w /\kappa}-1}. \label{omexdistribution} \ee
%\be \beta_{\w \w'} &=&  -\frac{e^{-i (\w+\w') v_H}}{2 \pi \kappa} \frac{\sqrt{\w \w'}}{\w + \w'} \, \left[-\frac{i}{\kappa} (\w+\w') \right]^{-i \w/\kappa} \Gamma \left( \frac{i \w}{\kappa} \right)  \;. \label{beta-mirror-3} \label{alpha-beta-mirror-3} \ee
It is easy to see that as $\xi \rightarrow 1$ that the spectra coincide. The high frequency approximation, $\w'\gg \w$, applied to Omex's distribution, Eq.~(\ref{omexdistribution}), gives the usual thermal result of Carlitz-Willey and late-time Davies-Fulling trajectories,
\be|\beta_{\w \w'}|^2_{T} = \frac{1}{2 \pi  \kappa  \w'}\frac{1}{ e^{2 \pi  \w/\kappa }-1}.\ee
One checks that the Domex distribution is thermal, $|\beta_{\w \w'\xi}|^2 \approx |\beta_{\w \w'}|^2_T$, by a series expansion that first approximates the distribution with a very fast end-state drifting mirror $\xi \approx 1$ and then applies the high-frequency approximation $\w'\gg \w$.  This approach explicitly decouples the late-time approximation from the high-frequency approximation.

\subsection{Consistency Check}
It is a fair claim that the total summation of the energies of each quanta should be equal to the integral over the energy flux:
\be \int_0^\infty \omega \la N_\omega \ra \d\omega = \int_{-\infty}^{\infty} F(u) \d u. \ee 
Therefore, the beta Bogoliubov particle results can be confirmed to be consistent with the stress-energy by computing the total energy, using Eq.~(\ref{particlecount}) and Eq.~(\ref{fluxtime}),
\be \label{quantasum} \int_0^\infty \omega \left[\int_0^\infty |\beta_{\omega\omega'}|^2 \d\omega'\right] \d\omega = \int_{-\infty}^{\infty} F(t) (1-\dot{z})\d t. \ee 
This consistency helps confirm the particles do indeed carry the energy.  
Explicitly, we have numerically confirmed the beta coefficients, Eq.~(\ref{particlebeta}), that for an observer to the right at $\mathscr{I}_R^+$, the total energy emitted is 
\be E_R = \int_0^\infty \int_0^\infty \w \cdot |\beta_{\w \w' \xi }|^2 \; \d\w \; \d\w' = \frac{\kappa  (3- \xi) \tanh ^{-1}(\xi )}{48 \pi  \xi ^2}-\frac{\kappa  (3+ 2\xi)}{48 \pi  (\xi^2 + \xi)},\ee
and for an observer to the left at $\mathscr{I}_L^+$, the total energy emitted is 
\be E_L = \int_0^\infty \int_0^\infty \w' \cdot |\beta_{\w \w' \xi }|^2 \; \d\w \; \d\w' = \frac{\kappa(3+ \xi) \tanh ^{-1}(-\xi )}{48 \pi  \xi ^2} -\frac{\kappa  (3-2\xi)}{48 \pi  (\xi^2 - \xi)},\ee%= -\frac{\kappa }{48\pi} \frac{\left(\xi  (2 \xi +3)+(\xi -3) (\xi +1) \tanh ^{-1}(\xi )\right)}{\xi ^2 (\xi +1)} \ee
in agreement with the analytical results of the stress-energy tensor of Section (\ref{sec:energy}).  We have found the largest relative numerical error here to be less than $10^{-11}$ using $\kappa =1$, and various values of $0< \xi <1$.   
%\be E_L = \int_0^\infty \int_0^\infty \w' * |\beta_{\w \w' \xi }|^2 \; d\w \; d\w' = \frac{\kappa  \left(4 \xi ^2-6 \xi +(\xi -1) (\xi +3) \log \left(\frac{2}{\xi +1}-1\right)\right)}{96 \pi  (\xi -1) \xi ^2},\ee
%\be E_L = \int_0^\infty \int_0^\infty \w' * |\beta_{\w \w' \xi }|^2 \; d\w \; d\w' = \frac{\kappa  \left(2 \xi-3\right)}{48 \pi  (\xi -1) \xi} +\frac{\kappa(\xi +3)}{96 \pi \xi ^2} \log \left(\frac{2}{\xi +1}-1\right),\ee%= -\frac{\kappa }{48\pi} \frac{\left(\xi  (2 \xi +3)+(\xi -3) (\xi +1) \tanh ^{-1}(\xi )\right)}{\xi ^2 (\xi +1)} \ee

\section{Discussions}\label{sec:conclusions} 
%\subsection{Horizonless Temperature}

In the Introduction we raised the following question: ``Is there a moving mirror in (1+1)-dimensions, satisfying unitarity in the sense allowed by the Bianchi-Smerlak criterion (namely, $S(u) \to \text{const.}$ as $u \to \pm\infty$), that has no acceleration horizon, produces finite amount of energy, and serves as a limiting case analog for Eddington-Finkelstein coordinate null shell gravitational collapse?'' We have answered this question in the affirmative, by constructing an exact mirror solution that satisfies these properties. 
Furthermore, we investigated both sides of the mirror trajectory, and found interesting features regarding negative entropy and negative energy flux.

The hallmark trait of the solution is the fact that it is an asymptotically coasting mirror which does not have an accelerating horizon, yet approaches arbitrarily close to thermal equilibrium.  Thermal radiation arises from a sufficiently fast final drifting speed. The ray-tracing function is identical to the spacetime matching condition of the black hole case in the limit that the mirror drifts to the speed of light.

The global approach to treating horizons tends to work well in fully equilibrium thermodynamics, especially so with \emph{a priori} non-dynamical assumptions (i.e. constant energy flux) \cite{Carlitz:1986nh}.  It is well-known that the non-equilibrium cases are not so easy to formulate using the traditional methods \cite{Raval:1996vt}.  \emph{A practical outcome of this paper has been to show how robust the traditional methods can be when the horizon is removed from the start.  Non-equilibrium dynamical conditions follow suit, however the system can still achieve equilibrium, for an arbitrary extended amount of time.} With a consistent scaling ($\kappa$ in Domex and Omex are the same scaled parameter relative to thermal emission), we have explicitly used the global geometric properties of the spacetime, and in the case of particle creation, only localized \textit{after} solving for the global beta Bogoliubov coefficients.   

%In this work we have shown that the appropriately generalized horizonless black mirror particle production and energy flux, evolved in time, exhibit manifest signatures of thermal radiation: \textit{horizonless temperature}.
%We have shown that Domex is the appropriately generalized drifting analog of Omex.  We have shown that Domex, while horizonless, has dynamic particle production and energy flux which exhibit manifest signatures of thermal radiation: \textit{horizonless temperature}.  
The new mirror was described in terms of its energy flux, total energy, entropy flux, correlations, and particle flux.  The temperature can be detected by asymptotic observers with particle detectors (the radiation demonstrates a Planckian distribution for a very fast final drift speed and the use of the high frequency approximation).  The evidence for thermality is strengthened further by the long-lived steady-state stress-energy tensor and the correlations which match the eternally thermal equilibrium mirror of Carlitz-Willey \cite{Carlitz:1986nh}. 
However, a few remarks are in order comparing our analysis to Carlitz-Willey's seperate 1987 trajectory \cite{Carlitz1987} which is not eternally thermal.  A notable similarity between this apparent horizon trajectory and the one presented here is the constant rate of particle emission during a finite period of time. We confirm the locally thermal state in both unitary moving mirror trajectories.  One notable difference is that their trajectory is in terms of an approximate ray-tracing function with a kink.\footnote{Carlitz and Willey comment that the kink can be smoothed out but it is not done because it is clear it would not affect the conclusions much.}  The trajectory in this paper is exact, $C^\infty$, for all-times, and is expressed as an explicit space-time trajectory function, Eq.~(\ref{trajectory}).  A consequence of this fact is that it so happens our trajectory does not come to rest at late times, while their trajectory requires the mirror eventually become stationary and consequently the entire remnant mass is radiated away to leave behind a flat region of spacetime.  In other words, Carlitz-Willey consider a ``meta-stable" or ``long-lived" remnant that slowly evaporates away, whereas our remnant is eternal.

Unlike the eternal thermal Carltiz-Willey mirror \cite{Carlitz:1986nh} or the black mirror \cite{Good:2016oey}, the Domex mirror gives rise to negative energy flux, and by the result of Bianchi-Smerlak \cite{Bianchi:2014vea,Bianchi:2014qua}, also to the non-monotonic mass loss of the any corresponding black hole.  Current efforts are being directed to explore the generalization of the tortoise coordinate from 
\be r^* \equiv r + 2 M \log \left(\frac{r}{2 M}-1\right)\ee
to
\be r^*(\xi) \equiv \frac{1+\xi}{1-\xi}r-\frac{4 M \xi }{1-\xi }-2 M \xi  W\left(\frac{2 e^{\frac{r/2M -1}{(1-\xi)/2}}}{1-\xi}\right), \label{tort} \ee
which can be evident from the generalization of the shock function of Eq.~(\ref{matchingfunction}) to Eq.~(\ref{matchingfunction2}), in the null-shell case which matches spacetimes outside and inside the shell.  For details on the null shell case, see e.g. Unruh (1976) \cite{Unruh:1976db}, Massar (1996) \cite{Massar:1996tx} or Fabbri (2005) \cite{Fabbri:2005mw}.  The generalization Eq.~(\ref{tort}) and a possible coupling between the parameters $\xi$ and $M$, may provide clues to understanding any corresponding black hole solution of Domex, and by necessity a different all-time collapse scenario.  It is understood\cite{Hawking2} that at very early times of gravitational collapse, the system cannot be described by the no-hair theorem.  Therefore it is appropriate to consider the type of modifications that can provide various early time approaches to a thermal distribution, particularly those modifications that can afford unitarity and finite evaporation energy. 
The modifications that can take into account energy conservation like those of the dilaton gravity models have had significant success as a laboratory for studying black hole evaporation. The physical problem in 1+1 dilaton gravity of the evaporating black hole and its modified emission extends to complete evaporation for the Russo, Susskind, and Thorlacius (RST) model \cite{Russo:1992ax} and to partial evaporation leaving a remnant for the Bose, Parker, and Peleg (BPP) model \cite{Bose:1995pz}.  The similarity of the Domex mirror to the BPP model is striking in several qualitative aspects: NEF emission as a thunderpop, a left over remnant, and finite total energy emission.  It is also interesting that the mass of the remnant in the BPP model is independent of the mass $M$ of the infalling matter, since with respect to the issue of energy conservation, there is no known physical analog for $M = 1/(4\kappa)$, the initial mass of the shockwave, in the mirror model.

Finally, we shall comment on the peculiar emission we find on the left side of the mirror trajectory, in particular, its possible relevance to the black hole correspondence (if one exists). We conjecture that the ``left emission'' corresponds to in-falling flux into the black hole. As such, its associated temperature and entropy might shed some light on the information paradox, since as we emphasized in the Introduction, unitarity is a property of the Hilbert space defined on the entire spacetime. In fact, such a ``left temperature'' in the context of black hole physics already exists in the literature, see e.g., \cite{1604.00465}. It should also be emphasized that the recent result in the literature \cite{1603.01964}, concerning the study of two-dimensional model of gravitational collapse, shows that a geodesic observer on the left side measures late time thermal radiation but zero flux. This result is drastically different from ours. It might be interesting to conduct a comparative study between our model with that of \cite{1603.01964}.

Ultimately, while Domex is elementary, it embraces several surprisingly interesting traits.  Since some of these traits are shared with more sophisticated systems, this solution may be a precursor for ensuing developments (the overt example being curved spacetime collapse).  On the other hand, this solution has exposed several explicit general attributes which are unanticipated and must be understood in order to claim a good grasp of the dynamics of the particle creation effect in non-thermal equilibrium.  

The outstanding advantage of this mirror solution is the exact expressions for quantities of interest.  Since one natural speculation is the direct applicability to a curved spacetime analog, we aim to examine this pertinent and interesting follow-up topic in a later manuscript with primary consideration to energy conservation of the black hole's modified evaporation emission, metric continuity across the shock boundary, and to the Bogolubov coefficients of specific dilaton gravity models.

%	One challenge we have not addressed is how does one decoupling the late time approximation from high frequency approximation? Another challenge is how can this be used to understand the NEF and the death gasp better?

%It is also worth noting that in a recent work of Abdolrahimi-Page \cite{AP}, the negative energy flux in the case of an asymptotically flat Schwarzschild black hole is shown to be extremely tiny. They found that the mass increase of the black hole is less than 0.09\% of the energy of a \emph{single} quantum of the energy of the Hawking temperature of said black hole at that time, and is therefore unlikely to be detectable, considering that such a signature would in addition be swamped by the noise of quantum fluctuations. Our model's aim is to purely study the detectability of the thermal plateau flux at infinity via particle production evolution.  

\section*{Acknowledgment}
MRRG thanks Paul Anderson, Xiong Chi and Charles Evans for insightful conversations, as well as the Julian Schwinger Foundation for its support under Grant 15-07-0000. YCO thanks the Department of Physics of Nazarbayev University for hospitality during his visit. YCO also thanks Nordita for support and a conducive environment for research, as well as the National Natural Science Foundation of China (NNSFC) for supporting his research in SJTU.

%%%%%%%%%%%%%%%%%%%%%%%%%%%%%%%%%%%%%%%%%%%%%%%%%%%%%%%%%%%%%%%%


\begin{thebibliography}{99}

\bibitem{Hawking2}
S. W. Hawking, ``Particle Creation by Black Holes'', {\changeurlcolor{vividviolet}\href{https://link.springer.com/article/10.1007/BF02345020}{Commun. Math. Phys. \textbf{43} (1975) 199}}.

\bibitem{Hawking3}
S. W. Hawking, ``Breakdown of Predictability in Gravitational Collapse'', {\changeurlcolor{vividviolet}\href{http://journals.aps.org/prd/abstract/10.1103/PhysRevD.14.2460}{Phys. Rev. D \textbf{14} (1976) 2460}}.

\bibitem{amps}
A. Almheiri, D. Marolf, J. Polchinski, J. Sully, ``Black Holes: Complementarity or Firewalls?'', {\changeurlcolor{vividviolet}\href{https://link.springer.com/article/10.1007\%2FJHEP02\%282013\%29062}{JHEP \textbf{1302} (2013) 062}}, \href{http://arxiv.org/abs/1207.3123}{[arXiv:1207.3123 [hep-th]]}.

\bibitem{apologia}
A. Almheiri, D. Marolf, J. Polchinski, D. Stanford, J. Sully,
``An Apologia for Firewalls'', {\changeurlcolor{vividviolet}\href{https://link.springer.com/article/10.1007\%2FJHEP09\%282013\%29018}{JHEP \textbf{1309} (2013) 018}}, \href{http://arxiv.org/abs/1304.6483}{[arXiv:1304.6483 [hep-th]]}.


\bibitem{sam}
S. L. Braunstein, S. Pirandola, K. \.Zyczkowski, ``Better Late than Never: Information Retrieval from Black Holes'', {\changeurlcolor{vividviolet}\href{http://journals.aps.org/prl/abstract/10.1103/PhysRevLett.110.101301}{Phys. Rev. Lett. \textbf{110} (2013) 101301}}, \href{http://arxiv.org/abs/0907.1190}{[arXiv:0907.1190 [quant-ph]]}.

\bibitem{naked}
P. Chen, Y. C. Ong, D. N. Page, M. Sasaki, D.-h. Yeom, ``Naked Black Hole Firewalls'',  {\changeurlcolor{vividviolet}\href{http://journals.aps.org/prl/abstract/10.1103/PhysRevLett.116.161304}{Phys. Rev. Lett. \textbf{116} (2016) 161304}}, \href{https://arxiv.org/abs/1511.05695}{[arXiv:1511.05695 [hep-th]]}.

\bibitem{huh}
Y. C. Ong, D.-h. Yeom, ``Black Hole Information Loss: Some Food for Thoughts'', to appear in the Proceedings
of the 2nd LeCosPA International Symposium, ``Everything About Gravity'', 2015, \href{https://arxiv.org/abs/1602.06600}{[arXiv:1602.06600 [hep-th]]}.

\bibitem{Landauer}
R. Landauer, ``The Physical Nature of Information'', {\changeurlcolor{vividviolet}\href{http://www.sciencedirect.com/science/article/pii/0375960196004537}{Phys. Lett. A \textbf{217} (1996) 188}}.

\bibitem{Adami}
C. Adami, ``The Physics of Information'', \href{https://arxiv.org/abs/quant-ph/0405005}{[arXiv:quant-ph/0405005]}.

\bibitem{Wolf}
D. Reeb, M. M. Wolf, ``An Improved Landauer Principle with Finite-Size Corrections'', {\changeurlcolor{vividviolet}\href{http://iopscience.iop.org/article/10.1088/1367-2630/16/10/103011/meta;jsessionid=5EE20E2C7E74743E5C39355C873099C2.c4.iopscience.cld.iop.org}{New J. Phys. \textbf{16} (2014) 103011}}, \href{https://arxiv.org/abs/1306.4352}{[arXiv:1306.4352 [quant-ph]]}.

\bibitem{0311049}
J. D. Bekenstein, ``Black Holes and Information Theory'', {\changeurlcolor{vividviolet}\href{http://www.tandfonline.com/doi/abs/10.1080/00107510310001632523}{Contemp. Phys. \textbf{45} (2003) 31}}, \href{https://arxiv.org/abs/quant-ph/0311049}{[arXiv:quant-ph/0311049]}.

\bibitem{0708.4025}
P. Hayden, J. Preskill, ``Black Holes as Mirrors: Quantum Information in Random Subsystems'', {\changeurlcolor{vividviolet}\href{http://iopscience.iop.org/article/10.1088/1126-6708/2007/09/120/meta}{JHEP 0709 (2007) 120}}, \href{https://arxiv.org/abs/0708.4025}{[arXiv:0708.4025 [hep-th]]}.

\bibitem{0808.2096}
Y. Sekino, L. Susskind, ``Fast Scramblers'', {\changeurlcolor{vividviolet}\href{http://iopscience.iop.org/article/10.1088/1126-6708/2008/10/065/meta}{JHEP \textbf{10} (2008) 065}}, \href{https://arxiv.org/abs/0808.2096}{[arXiv:0808.2096 [hep-th]]}.

\bibitem{1111.6580}
N. Lashkari, D. Stanford, M. Hastings, T. Osborne, P. Hayden, ``Towards the Fast Scrambling Conjecture'', {\changeurlcolor{vividviolet}\href{https://link.springer.com/article/10.1007\%2FJHEP04\%282013\%29022}{JHEP \textbf{04} (2013) 022}}, \href{https://arxiv.org/abs/1111.6580}{[arXiv:1111.6580 [hep-th]]}.

\bibitem{1301.4504}
D. Harlow, P. Hayden, ``Quantum Computation vs. Firewalls'', {\changeurlcolor{vividviolet}\href{https://link.springer.com/article/10.1007\%2FJHEP06\%282013\%29085}{JHEP \textbf{06} (2013) 085}}, \href{https://arxiv.org/abs/1301.4504}{[arXiv:1301.4504 [hep-th]]}.

\bibitem{1301.4505}
L. Susskind, ``Black Hole Complementarity and the Harlow-Hayden Conjecture'', \href{https://arxiv.org/abs/1301.4505}{[arXiv:1301.4505 [hep-th]]}.


\bibitem{1402.5674}
L. Susskind, ``Computational Complexity and Black Hole Horizons'', \href{https://arxiv.org/abs/1402.5674}{[arXiv:1402.5674 [hep-th]]}.

\bibitem{page1}
D. N. Page, ``Average Entropy of a Subsystem'', {\changeurlcolor{vividviolet}\href{http://journals.aps.org/prl/abstract/10.1103/PhysRevLett.71.1291}{Phys. Rev. Lett. \textbf{71} (1993) 1291}}, \href{http://arxiv.org/abs/gr-qc/9305007}{[arXiv:gr-qc/9305007]}.

\bibitem{page1b}
D. N. Page, ``Information in Black Hole Radiation'', {\changeurlcolor{vividviolet}\href{http://journals.aps.org/prl/abstract/10.1103/PhysRevLett.71.3743}{Phys. Rev. Lett. \textbf{71} (1993) 3743}}, \href{http://arxiv.org/abs/hep-th/9306083}{[arXiv:hep-th/9306083]}.

\bibitem{page2}
D. N. Page, ``Time Dependence of Hawking Radiation Entropy'', {\changeurlcolor{vividviolet}\href{http://iopscience.iop.org/article/10.1088/1475-7516/2013/09/028/meta}{JCAP \textbf{1309} (2013) 028}} , \href{http://arxiv.org/abs/1301.4995}{[arXiv:1301.4995 [hep-th]]}.

\bibitem{0901.3156}
S. Hossenfelder, L. Smolin, ``Conservative Solutions to the Black Hole Information Problem'', 	{\changeurlcolor{vividviolet}\href{http://journals.aps.org/prd/abstract/10.1103/PhysRevD.81.064009}{Phys. Rev. D \textbf{81} (2010) 064009}}, \href{https://arxiv.org/abs/0901.3156}{[arXiv:0901.3156 [gr-qc]9]}.

\bibitem{9403137}
T. M. Fiola, J. Preskill, A. Strominger, S. P. Trivedi, ``Black Hole Thermodynamics and Information Loss in Two Dimensions'', {\changeurlcolor{vividviolet}\href{http://journals.aps.org/prd/abstract/10.1103/PhysRevD.50.3987}{Phys. Rev. D \textbf{50} (1994) 3987}}, \href{https://arxiv.org/abs/hep-th/9403137}{[arXiv:hep-th/9403137]}.

\bibitem{Bianchi:2014vea} 
  E.~Bianchi, M.~Smerlak,
  ``Last Gasp of a Black Hole: Unitary Evaporation Implies Non-Monotonic Mass Loss'',
  {\changeurlcolor{vividviolet}\href{https://link.springer.com/article/10.1007\%2Fs10714-014-1809-9}{Gen.\ Rel.\ Grav.\  {\bf 46} (2014) 1809}},
  \href{http://arxiv.org/abs/1405.5235}{[arXiv:1405.5235 [gr-qc]]}.

\bibitem{Bianchi:2014qua} 
  E.~Bianchi, M.~Smerlak,
  ``Entanglement Entropy and Negative Energy in Two Dimensions'',
  {\changeurlcolor{vividviolet}\href{http://journals.aps.org/prd/abstract/10.1103/PhysRevD.90.041904}{Phys.\ Rev.\ D {\bf 90} (2014) 041904}},
  \href{http://arxiv.org/abs/1404.0602}{[arXiv:1404.0602 [gr-qc]]}.
	
	\bibitem{1409.0144}
	E. Bianchi, T. De Lorenzo, M. Smerlak,
	``Entanglement Entropy Production in Gravitational Collapse: Covariant Regularization and Solvable Models'', {\changeurlcolor{vividviolet}\href{https://link.springer.com/article/10.1007\%2FJHEP06\%\282015\%29180}{JHEP \textbf{06} (2015) 180}}, \href{https://arxiv.org/abs/1409.0144}{[arXiv:1409.0144 [hep-th]]}.
	
	
		
		
	\bibitem{AP}
	S. Abdolrahimi, D. N. Page, ``Hawking Radiation Energy and Entropy from a Bianchi-Smerlak Semiclassical Black Hole'', {\changeurlcolor{vividviolet}\href{http://journals.aps.org/prd/abstract/10.1103/PhysRevD.92.083005}{Phys. Rev. D \textbf{92} (2015) 083005}}, \href{http://arxiv.org/abs/1506.01018}{[arXiv:1506.01018 [hep-th]]}.
	
	\bibitem{9508027}
	S. Bose, L. Parker, Y. Peleg, ``Hawking Radiation and Unitary Evolution'', 	{\changeurlcolor{vividviolet}\href{http://journals.aps.org/prl/abstract/10.1103/PhysRevLett.76.861}{Phys. Rev. Lett. \textbf{76} (1996) 861}}, \href{https://arxiv.org/abs/gr-qc/9508027v2}{[arXiv:gr-qc/9508027]}.
	
	\bibitem{Good:2015nja} 
  M.~R.~R.~Good, Y.~C.~Ong,
  ``Signatures of Energy Flux in Particle Production: A Black Hole Birth Cry and Death Gasp'',
  {\changeurlcolor{vividviolet}\href{https://link.springer.com/article/10.1007\%2FJHEP07\%282015\%29145}{JHEP {\bf 1507} (2015)145}},
  \href{https://arxiv.org/abs/1506.08072}{[arXiv:1506.08072 [gr-qc]]}.
	
	\bibitem{9403108v1}
C. Holzhey, F. Larsen, F. Wilczek, ``Geometric and Renormalized Entropy in Conformal Field Theory'', {\changeurlcolor{vividviolet}\href{http://www.sciencedirect.com/science/article/pii/0550321394904022}{Nucl. Phys. B \textbf{424} (1994) 443}}, \href{https://arxiv.org/abs/hep-th/9403108v1}{[arXiv:hep-th/9403108]}.

		\bibitem{aharonov}
	Y. Aharonov, A. Casher, S. Nussinov, ``The Unitarity Puzzle and Planck Mass Stable Particles'', {\changeurlcolor{vividviolet}\href{http://www.sciencedirect.com/science/article/pii/0370269387913207}{Phys. Lett. B \textbf{191} (1987) 51}}.
	
	\bibitem{Chen:2014jwq} 
  P.~Chen, Y.~C.~Ong, D.-h.~Yeom,
  ``Black Hole Remnants and the Information Loss Paradox'',
	{\changeurlcolor{vividviolet}\href{http://www.sciencedirect.com/science/article/pii/S0370157315004391}{Physics Reports (2015) 1}},
  \href{http://arxiv.org/abs/1412.8366}{[arXiv:1412.8366 [gr-qc]]}.
	
	\bibitem{1411.2854}
	M. Christodoulou, C. Rovelli,
	``How Big is a Black Hole?'', {\changeurlcolor{vividviolet}\href{http://journals.aps.org/prd/abstract/10.1103/PhysRevD.91.064046}{Phys. Rev. D \textbf{91} (2015) 064046}},  \href{https://arxiv.org/abs/1411.2854}{[arXiv:1411.2854 [gr-qc]]}.
	
	%\cite{Bhaumik:2016sav}
\bibitem{Bhaumik:2016sav} 
  N.~Bhaumik and B.~R.~Majhi,
  ``Interior volumes of extremal and ($1+D$) dimensional Schwarzschild black holes,'' \href{https://arxiv.org/abs/1607.03704}{[arXiv:1607.03704 [gr-qc]]}.
  %%CITATION = ARXIV:1607.03704;%%
	
		\bibitem{1503.08245}
	Y. C. Ong, ``The Persistence of the Large Volumes in Black Holes'', {\changeurlcolor{vividviolet}\href{https://link.springer.com/article/10.1007\%2Fs10714-015-1929-x}{Gen. Relativ. Gravit. \textbf{47} (2015) 88}}, \href{https://arxiv.org/abs/1503.08245}{[arXiv:1503.08245 [gr-qc]]}.

	
	\bibitem{1602.04395}
	Y. C. Ong, ``Black Hole: The Interior Spacetime'',  to appear in the Proceedings
of the 2nd LeCosPA International Symposium, ``Everything About Gravity'', 2015, \href{https://arxiv.org/abs/1602.04395}{[arXiv:1602.04395 [gr-qc]]}.

\bibitem{1604.07222}
M. Christodoulou, T. De Lorenzo, ``On the Volume Inside Old Black Holes'', \href{https://arxiv.org/abs/1604.07222}{[arXiv:1604.07222 [gr-qc]]}.
		
	\bibitem{1510.02182}
	B. C. Zhang, ``Entropy in the Interior of a Black Hole and Thermodynamics'', 	{\changeurlcolor{vividviolet}\href{http://journals.aps.org/prd/abstract/10.1103/PhysRevD.92.081501}{Phys. Rev. D \textbf{92} (2015) 081501(R)}}, \href{https://arxiv.org/abs/1510.02182}{[arXiv:1510.02182 [gr-qc]]}.



\bibitem{Davies:1976hi} 
  S.~A.~Fulling, P.~C.~W.~Davies,
  ``Radiation from a Moving Mirror in Two-Dimensional Space-Time Conformal Anomaly'',
  {\changeurlcolor{vividviolet}\href{http://rspa.royalsocietypublishing.org/content/348/1654/393}{Proc.\ Roy.\ Soc.\ Lond.\ A {\bf 348} (1976) 393}}.


\bibitem{Davies:1977yv} 
  P.~C.~W.~Davies, S.~A.~Fulling,
  ``Radiation from Moving Mirrors and from Black Holes'',
  {\changeurlcolor{vividviolet}\href{http://rspa.royalsocietypublishing.org/content/356/1685/237}{Proc.\ Roy.\ Soc.\ Lond.\ A {\bf 356}  (1977) 237}}.
	
	\bibitem{D0}
	  P.~C.~W.~Davies, ``Thermodynamics of Black Holes'', {\changeurlcolor{vividviolet}\href{http://iopscience.iop.org/article/10.1088/0034-4885/41/8/004/pdf}{Rep. Prog. Phys. \textbf{41} (1978) 1313}}.

	\bibitem{moore}
G.~T.~Moore,
``Quantum Theory of the Electromagnetic Field in a Variable-Length One-Dimensional Cavity'',
{\changeurlcolor{vividviolet}\href{http://scitation.aip.org/content/aip/journal/jmp/11/9/10.1063/1.1665432}{J.\ Math.\ Phys. \textbf{11} (1970) 2679}}.

\bibitem{DeWitt:1975ys} 
  B.~S.~DeWitt,
  ``Quantum Field Theory in Curved Space-Time'', 
  {\changeurlcolor{vividviolet}\href{http://www.sciencedirect.com/science/article/pii/0370157375900514}{Phys.\ Rept.\  {\bf 19} (1975) 295}}.

	\bibitem{Silva:2016wgo} 
  J.~D.~L.~Silva, A.~N.~Braga, D.~T.~Alves,
  ``Dynamical Casimir Effect with $\delta-\delta^{\prime}$ Mirrors'', 
  \href{https://arxiv.org/abs/1601.06109}{[arXiv:1601.06109 [quant-ph]]}.


	




	
	
%\bibitem{sabine}
%S. Hossenfelder, ``Disentangling the Black Hole Vacuum'', 	Phys. Rev. D \textbf{91} (2015) 044015, \href{http://arxiv.org/abs/1401.0288}{[arXiv:1401.0288 [hep-th]]}.

%\bibitem{1409.7754}
%M. Visser, ``Thermality of the Hawking Flux'', \href{http://arxiv.org/abs/1409.7754}{[arXiv:1409.7754 [gr-qc]]}.



\bibitem{Stargen:2016euf} 
  D.~J.~Stargen, D.~A.~Kothawala, L.~Sriramkumar,
  ``Moving Mirrors and the Fluctuation-Dissipation Theorem'',
{\changeurlcolor{vividviolet}\href{http://journals.aps.org/prd/abstract/10.1103/PhysRevD.94.025040}{Phys. Rev. D \textbf{94} (2016) 025040}},
  \href{https://arxiv.org/abs/1602.02526}{[arXiv:1602.02526 [hep-th]]}.

	

\bibitem{Wang:2015axa} 
  Q.~Wang, W.~G.~Unruh,
  ``Mirror Moving in Quantum Vacuum of a Massive Scalar Field'', 
 {\changeurlcolor{vividviolet}\href{http://journals.aps.org/prd/abstract/10.1103/PhysRevD.92.063520}{Phys.\ Rev.\ D {\bf 92} (2015) 063520}},
  \href{https://arxiv.org/abs/1506.05531}{[arXiv:1506.05531 [gr-qc]]}.

\bibitem{Hotta:2015yla} 
  M.~Hotta, R.~Schützhold, W.~G.~Unruh,
  ``Partner Particles for Moving Mirror Radiation and Black Hole Evaporation'', 
 {\changeurlcolor{vividviolet}\href{http://journals.aps.org/prd/abstract/10.1103/PhysRevD.91.124060}{Phys.\ Rev.\ D {\bf 91} (2015) 124060}},
  \href{https://arxiv.org/abs/1503.06109}{[arXiv:1503.06109 [gr-qc]]}.


\bibitem{Yeh:2013mca} 
  C.~P.~Yeh, J.~T.~Hsiang, D.~S.~Lee,
  ``Holographic Approach to Nonequilibrium Dynamics of Moving Mirrors Coupled to Quantum Critical Theories'',
  {\changeurlcolor{vividviolet}\href{http://journals.aps.org/prd/abstract/10.1103/PhysRevD.89.066007}{Phys.\ Rev.\ D {\bf 89} (2014) 066007}},
 \href{https://arxiv.org/abs/1310.8416}{[arXiv:1310.8416 [hep-th]]}.

	

\bibitem{Nicolaevici:2014ela} 
  N.~Nicolaevici,
  ``Unruh Effect Without Rindler Horizon,''
  {\changeurlcolor{vividviolet}\href{http://iopscience.iop.org/article/10.1088/0264-9381/32/4/045013/meta;jsessionid=B9BDDEF29239FE5E9B3C70AFC3438337.c2.iopscience.cld.iop.org}{Class.\ Quant.\ Grav.\  {\bf 32} (2015) 045013}},  
  \href{https://arxiv.org/abs/1501.00119}{[arXiv:1501.00119 [hep-th]]}.
	
	\bibitem{Silva:2011fq} 
  H.~O.~Silva, C.~Farina,
  ``A Simple Model for the Dynamical Casimir Effect for a Static Mirror with Time-Dependent Properties'', 
  {\changeurlcolor{vividviolet}\href{http://journals.aps.org/prd/abstract/10.1103/PhysRevD.84.045003}{Phys.\ Rev.\ D {\bf 84} (2011) 045003}}
  \href{https://arxiv.org/abs/1102.2238}{[arXiv:1102.2238 [hep-th]]}.


\bibitem{Hotta:2015huj} 
  M.~Hotta, A.~Sugita,
  ``The Fall of Black Hole Firewall: Natural Nonmaximal Entanglement for Page Curve'', 
  {\changeurlcolor{vividviolet}\href{http://ptep.oxfordjournals.org/content/2015/12/123B04}{PTEP {\bf 2015} (2015) 123B04}}, 
  \href{https://arxiv.org/abs/1505.05870}{[arXiv:1505.05870 [gr-qc]]}.
  


\bibitem{Haro:2008zza} 
  J.~Haro, E.~Elizalde,
  ``Black Hole Collapse Simulated by Vacuum Fluctuations with a Moving Semitransparent Mirror'',
  {\changeurlcolor{vividviolet}\href{http://journals.aps.org/prd/abstract/10.1103/PhysRevD.77.045011}{Phys.\ Rev.\ D {\bf 77} (2008) 045011}},
	\href{https://arxiv.org/abs/0712.4141}{[arXiv:0712.4141 [quant-ph]]}
	
 
	\bibitem{Alves:2010zza} 
  D.~T.~Alves, E.~R.~Granhen, H.~O.~Silva, M.~G.~Lima,
  ``Quantum Radiation Force on the Moving Mirror of a Cavity, with Dirichlet and Neumann Boundary Conditions for a Vacuum, Finite Temperature, and a Coherent State'',
  {\changeurlcolor{vividviolet}\href{http://journals.aps.org/prd/abstract/10.1103/PhysRevD.81.025016}{Phys.\ Rev.\ D {\bf 81} (2010) 025016}}.


\bibitem{Good:2012cp} 
  M.~R.~R.~Good, ``On Spin-Statistics and Bogoliubov Transformations in Flat Spacetime With Acceleration Conditions'',
  {\changeurlcolor{vividviolet}\href{http://www.worldscientific.com/doi/abs/10.1142/S0217751X13500085}{Int.\ J.\ Mod.\ Phys.\ A {\bf 28} (2013)1350008}},
  \href{https://arxiv.org/abs/1205.0881}{[arXiv:1205.0881 [gr-qc]]}.

\bibitem{Chen:2015bcg} 
  P.~Chen, G.~Mourou,
  ``Accelerating Plasma Mirrors to Investigate Black Hole Information Loss Paradox'',
  \href{https://arxiv.org/abs/1512.04064}{[arXiv:1512.04064 [gr-qc]]}.
	
	\bibitem{Wilson:2011} 
  C.~M.~Wilson, G.~Johansson, A.~Pourkabirian, J.~R.~Johansson, T.~Duty, F.~Nori, P.~Delsing,
  ``Observation of the Dynamical Casimir Effect in a Superconducting Circuit'', {\changeurlcolor{vividviolet}\href{http://www.nature.com/nature/journal/v479/n7373/full/nature10561.html}{Nature \textbf{479} (2011) 376}}, \href{https://arxiv.org/abs/1105.4714}{[arXiv:1105.4714 [quant-ph]]}.

\bibitem{Lahteenmaki:2013mda} 
  P.~Lahteenmaki, G.~S.~Paraoanu, J.~Hassel, P.~J.~Hakonen,
  ``Dynamical Casimir Effect in a Josephson Metamaterial'', 
{ \changeurlcolor{vividviolet}\href{http://www.pnas.org/content/110/11/4234}{Proc. Natl. Acad. Sci. U.S.A. {\bf 110} (2013) 4234}}, \href{https://arxiv.org/abs/1111.5608}{[arXiv:1111.5608 [cond-mat.mes-hall]]}.

\bibitem{Good:2016oey} 
  M.~R.~R.~Good, P.~R.~Anderson, C.~R.~Evans,
  ``Mirror Reflections of a Black Hole'', {\changeurlcolor{vividviolet}\href{http://journals.aps.org/prd/abstract/10.1103/PhysRevD.94.065010}{Phys. Rev. D \textbf{94} (2016) 065010}}, 
  \href{https://arxiv.org/abs/1605.06635}{[arXiv:1605.06635 [gr-qc]]}.

	
\bibitem{paper1} 
P. R. Anderson,  M.~R.~R.~Good, C.~R.~Evans, ``Black Hole - Moving Mirror I: An Exact Correspondence'', to appear in the Proceedings of the 14th Marcel Grossmann Meeting, 2015,
\href{https://arxiv.org/abs/1507.03489}{[arXiv:1507.03489 [gr-qc]]}.

\bibitem{paper2} 
  M.~R.~R.~Good, P.~R.~Anderson, C.~R.~Evans,
  ``Black Hole - Moving Mirror II: Particle Creation'', to appear in the Proceedings of the 14th Marcel Grossmann Meeting, 2015,
  \href{https://arxiv.org/abs/1507.05048}{[arXiv:1507.05048 [gr-qc]]}.
 
	
%\cite{Good:2016bsq}
\bibitem{Good:2016bsq} 
  M.~R.~R.~Good,
  ``Reflections on a Black Mirror'', to appear in the Proceedings of the 2nd LeCosPA International Symposium,``Everything About Gravity", 2015,
  \href{https://arxiv.org/abs/1602.00683}{[arXiv:1602.00683 [gr-qc]]}.


\bibitem{Carlitz:1986nh} 
  R.~D.~Carlitz, R.~S.~Willey,
  ``Reflections On Moving Mirrors'', 
  {\changeurlcolor{vividviolet} \href{http://journals.aps.org/prd/abstract/10.1103/PhysRevD.36.2327}{Phys.\ Rev.\ D {\bf 36} (1987) 2327}}.


	\bibitem{Walker:1982} 
  W.~R.~Walker, P.~C.~W.~Davies,
  ``An Exactly Soluble Moving-Mirror Problem'',
  {\changeurlcolor{vividviolet} \href{http://iopscience.iop.org/article/10.1088/0305-4470/15/9/008}{J. Phys.\ A: Math. Gen. {\bf 15} (1982) L477}}.
	
	\bibitem{Good:2013lca} 
  M.~R.~R.~Good, P.~R.~Anderson, C.~R.~Evans,
  ``Time Dependence of Particle Creation from Accelerating Mirrors'', 
  {\changeurlcolor{vividviolet} \href{http://journals.aps.org/prd/abstract/10.1103/PhysRevD.88.025023}{Phys.\ Rev.\ D {\bf 88} (2013) 025023}},
  \href{http://arxiv.org/abs/1303.6756}{[arXiv:1303.6756 [gr-qc]]}.
	
	\bibitem{hawkingellis}
	S. W. Hawking, G. F. R. Ellis, \emph{The Large Scale Structure of Space-Time}, Cambridge Monographs on Mathematical Physics, Cambridge University Press, 1973 (reprinted 1994).
	
	\bibitem{cardy}
	J. Cardy, ``Operator Content of Two-Dimensional Conformal Invariant Theory'', {\changeurlcolor{vividviolet} \href{http://www.sciencedirect.com/science/article/pii/0550321386905523}{Nucl. Phys. B \textbf{270} (1986) 186}}.
	
	\bibitem{Raval:1996vt} 
  A.~Raval, B.~L.~Hu, D.~Koks,
  ``Near-Thermal Radiation in Detectors, Mirrors and Black Holes: A Stochastic Approach'', 
  {\changeurlcolor{vividviolet}\href{http://journals.aps.org/prd/abstract/10.1103/PhysRevD.55.4795}{Phys.\ Rev.\ D {\bf 55} (1997) 4795}},
  \href{https://arxiv.org/abs/gr-qc/9606074}{[arXiv:gr-qc/9606074]}.

 
\bibitem{1011.5593}
C. Barcelo, S. Liberati, S. Sonego, M. Visser, ``Minimal Conditions for the Existence of a Hawking-like Flux'', {\changeurlcolor{vividviolet}\href{http://journals.aps.org/prd/abstract/10.1103/PhysRevD.83.041501}{Phys. Rev. D \textbf{83} (2011) 041501}}, \href{https://arxiv.org/abs/1011.5593}{[arXiv:1011.5593 [gr-qc]]}.

\bibitem{Kawai:2013mda} 
  H.~Kawai, Y.~Matsuo and Y.~Yokokura,
  ``A Self-consistent Model of the Black Hole Evaporation,''
  Int.\ J.\ Mod.\ Phys.\ A {\bf 28}, 1350050 (2013)
  %doi:10.1142/S0217751X13500504
	\href{https://arxiv.org/abs/1302.4733}{[arXiv:1302.4733 [hep-th]]}.
   %%CITATION = doi:10.1142/S0217751X13500504;%%

\bibitem{1510.07157}
P.-M. Ho, ``The Absence of Horizon in Black-Hole Formation'',  {\changeurlcolor{vividviolet}\href{http://www.sciencedirect.com/science/article/pii/S0550321316301274}{Nucl. Phys. B \textbf{909} (2016) 394}}, \href{https://arxiv.org/abs/1510.07157}{[arXiv:1510.07157 [hep-th]]}.


	
\bibitem{kodama}
H. Kodama, ``Conserved Energy Flux for the Spherically Symmetric System and the Backreaction Problem in the Black Hole Evaporation'', {\changeurlcolor{vividviolet}\href{http://ptp.oxfordjournals.org/content/63/4/1217.abstract}{Prog. Theor. Phys.\textbf{63} (1980) 1217}}. 
	
\bibitem{hiscock1}
W. A. Hiscock, ``Models of Evaporating Black Holes. I'', {\changeurlcolor{vividviolet}\href{http://journals.aps.org/prd/abstract/10.1103/PhysRevD.23.2813}{Phys. Rev. D \textbf{23} (1981) 2813}}.
	
\bibitem{hiscock2}
W. A. Hiscock, ``Models of Evaporating Black Holes. II. Effects of the Outgoing Created Radiation'', {\changeurlcolor{vividviolet}\href{http://journals.aps.org/prd/abstract/10.1103/PhysRevD.23.2823}{Phys. Rev. D \textbf{23} (1981) 2823}}
	
\bibitem{kuroda}
Y. Kuroda, ``Vaidya Spacetime as an Evaporating Black Hole'', {\changeurlcolor{vividviolet}\href{http://ptp.oxfordjournals.org/content/71/6/1422.abstract}{Prog. Theor. Phys. \textbf{71} (1984) 1422}}.
	
\bibitem{Wilczek:1993jn} 
  F.~Wilczek,
  ``Quantum Purity at a Small Price: Easing a Black Hole Paradox'', 
  In *Houston 1992, Proceedings, Black Holes, Membranes, Wormholes and Superstrings* 1-21, and Inst. Adv. Stud. Princeton - IASSNS-HEP-93-012 (93/02,rec.Mar.) 19 p. (306377),
 \href{http://arxiv.org/abs/hep-th/9302096}{[arXiv: hep-th/9302096]}.


\bibitem{Walker:1984vj} 
  W.~R.~Walker,
  ``Particle and Energy Creation by Moving Mirrors'',
  {\changeurlcolor{vividviolet}\href{http://journals.aps.org/prd/abstract/10.1103/PhysRevD.31.767}{Phys.\ Rev.\ D {\bf 31} (1985) 767}}.
	
\bibitem{Ford:2004ba} 
  L.~H.~Ford, T.~A.~Roman,
  ``Energy Flux Correlations and Moving Mirrors'',
  {\changeurlcolor{vividviolet}\href{http://journals.aps.org/prd/abstract/10.1103/PhysRevD.70.125008}{Phys.\ Rev.\ D {\bf 70} (2004) 125008}},
  \href{https://arxiv.org/abs/gr-qc/0409093}{[arXiv: gr-qc/0409093]}.



	
\bibitem{Unruh:1976db} 
  W.~G.~Unruh,
  ``Notes on Black Hole Evaporation'',
  {\changeurlcolor{vividviolet}\href{http://journals.aps.org/prd/abstract/10.1103/PhysRevD.14.870}{Phys.\ Rev.\ D {\bf 14}  (1976) 870}}.

	
\bibitem{Massar:1996tx} 
  S.~Massar, R.~Parentani,
  ``From Vacuum Fluctuations to Radiation. 2. Black Holes'',
  {\changeurlcolor{vividviolet}\href{http://journals.aps.org/prd/abstract/10.1103/PhysRevD.54.7444}{Phys.\ Rev.\ D {\bf 54} (1996) 7444}}.


\bibitem{Fabbri:2005mw} 
  A.~Fabbri, J.~Navarro-Salas,
  \emph{Modeling Black Hole Evaporation},  
  London, UK: Imp. Coll. Pr., 2005.
	%%%%%%%%%%%%%%%%%%%%%%%%%%%%%%%%%%%%%%%%%%%%%%%%%%%%%%  <--- SEPARATION LINE: ABOVE THIS LINE REFs ARE IN ORDER
	%\cite{Carlitz1987}
\bibitem{Carlitz1987} 
  R.~D.~Carlitz and R.~S.~Willey,
  ``The Lifetime of a Black Hole,''
  Phys.\ Rev.\ D {\bf 36}, 2336 (1987).
  %doi:10.1103/PhysRevD.36.2336
  %%CITATION = doi:10.1103/PhysRevD.36.2336;%%
  %48 citations counted in INSPIRE as of 11 Jan 2017


%\cite{Russo:1992ax}
\bibitem{Russo:1992ax} 
  J.~G.~Russo, L.~Susskind and L.~Thorlacius,
  ``The Endpoint of Hawking radiation,''
  Phys.\ Rev.\ D {\bf 46}, 3444 (1992)
  %doi:10.1103/PhysRevD.46.3444
  [hep-th/9206070].
  %%CITATION = doi:10.1103/PhysRevD.46.3444;%%
  %247 citations counted in INSPIRE as of 11 Jan 2017

%\cite{Bose:1995pz}
\bibitem{Bose:1995pz} 
  S.~Bose, L.~Parker and Y.~Peleg,
  ``Semiinfinite throat as the end state geometry of two-dimensional black hole evaporation,''
  Phys.\ Rev.\ D {\bf 52}, 3512 (1995)
  %doi:10.1103/PhysRevD.52.3512
  [hep-th/9502098].
  %%CITATION = doi:10.1103/PhysRevD.52.3512;%%
  %53 citations counted in INSPIRE as of 11 Jan 2017
	

\bibitem{1604.00465}
Wontae Kim, ``Origin of Hawking Radiation: Firewall or Atmosphere?'', \href{https://arxiv.org/abs/1604.00465v3}{[arXiv:1604.00465 [hep-th]]}.
  
\bibitem{1603.01964}
Kinjalk Lochan, Sumanta Chakraborty, T. Padmanabhan, ``Dynamic Realization of the Unruh Effect for a Geodesic Observer'', \href{https://arxiv.org/abs/1603.01964}{[	arXiv:1603.01964 [gr-qc]]}.


		

\end{thebibliography}
\end{document}